\documentclass[reprint,twocolumn,prl,superscriptaddress]{revtex4-1}
\usepackage{graphicx,graphics,epsfig}
\usepackage{amsmath, amssymb,amsfonts} 
\usepackage{bm,times,xspace}  
\usepackage{color}
\usepackage[colorlinks, citecolor=blue, linkcolor=red, citecolor=blue, urlcolor=blue]{hyperref}
\usepackage{float}
\begin{document}

\title{Elastic Valley Spin Controlled Chiral Coupling in Topological Valley Phononic Crystals}
\author{Jinfeng Zhao}
\thanks{
These authors contributed equally: Jinfeng Zhao, Chenwen Yang, Weitao Yuan}
\affiliation{School of Aerospace Engineering and Applied Mechanics, Tongji University, 100 Zhangwu Road, Shanghai 200092, China}

\author{Chenwen Yang}
\thanks{
These authors contributed equally: Jinfeng Zhao, Chenwen Yang, Weitao Yuan}
\affiliation{Center for Phononics and Thermal Energy Science, China-EU Joint Lab on Nanophononics, Shanghai Key Laboratory of Special Artificial Microstructure Materials and Technology, School of Physics Science and Engineering, Tongji University, Shanghai 200092, China}

\author{Weitao Yuan}
\thanks{
These authors contributed equally: Jinfeng Zhao, Chenwen Yang, Weitao Yuan}
\affiliation{School of Aerospace Engineering and Applied Mechanics, Tongji University, 100 Zhangwu Road, Shanghai 200092, China}

\author{Danmei Zhang}
\affiliation{Center for Phononics and Thermal Energy Science, China-EU Joint Lab on Nanophononics, Shanghai Key Laboratory of Special Artificial Microstructure Materials and Technology, School of Physics Science and Engineering, Tongji University, Shanghai 200092, China}

\author{Yang Long}
\affiliation{Center for Phononics and Thermal Energy Science, China-EU Joint Lab on Nanophononics, Shanghai Key Laboratory of Special Artificial Microstructure Materials and Technology, School of Physics Science and Engineering, Tongji University, Shanghai 200092, China}

\author{Yongdong Pan}
\affiliation{School of Aerospace Engineering and Applied Mechanics, Tongji University, 100 Zhangwu Road, Shanghai 200092, China}

\author{Hong Chen}
\affiliation{Center for Phononics and Thermal Energy Science, China-EU Joint Lab on Nanophononics, Shanghai Key Laboratory of Special Artificial Microstructure Materials and Technology, School of Physics Science and Engineering, Tongji University, Shanghai 200092, China}

\author{Zheng Zhong}
\email{
Corresponding address, Zheng Zhong: zhongk@tongji.edu.cn}
\affiliation{School of Aerospace Engineering and Applied Mechanics, Tongji University, 100 Zhangwu Road, Shanghai 200092, China}

\author{Jie Ren}
\email{
Corresponding address, Jie Ren: Xonics@tongji.edu.cn}
\affiliation{Center for Phononics and Thermal Energy Science, China-EU Joint Lab on Nanophononics, Shanghai Key Laboratory of Special Artificial Microstructure Materials and Technology, School of Physics Science and Engineering, Tongji University, Shanghai 200092, China}

\begin{abstract}
Distinct from the phononic valley pseudo-spin, the real physical spin of elastic waves adds a novel tool-kit capable of envisaging the valley-spin physics of topological valley phononic crystals from a local viewpoint. 
Here, we report the observation of local elastic valley spin as well as the hidden elastic spin-valley locking mechanism overlooked before. 
We demonstrate that the selective one-way routing of valley phonon states along the topological interface can be reversed by imposing the elastic spin meta-source at different interface locations with opposite valley-spin correspondence. We unveil the physical mechanism of selective directionality as the elastic spin controlled chiral coupling of valley phonon states, through both analytical theory and  experimental measurement of the opposite local elastic spin density at different interface locations for different transport directions.
The elastic spin of valley topological edge phonons can be extended to other topological states and offers new tool to explore topological metamaterials. 
\end{abstract}

\maketitle

The valley in the momentum space that provides an extra degree of freedom (DOF) to manipulate particles has attracted tremendous interests in the field of topological materials in parallel to charge and spin ~\cite{Valleytronics,lundeberg2014harnessing,science.1250140}. Beside electron systems, the valley DOF has been 
successfully extended to classical wave systems like optical~\cite{liu2016pseudospin,opt_valley_hongchen_2015,opt_NM_2017_ZhangX,opt_NP_2018_ZhangBL,opt_PRL_2018_Rechtsman,PhysRevLett.122.123903}, acoustic~\cite{acstc_NP_2017_LZY,acstc_NC_2018_ChenYF,PhysRevLett.126.156401,acstc_NC_2020_TonyJH,PhysRevApplied.15.024019,adma.201803229} and elastic~\cite{elstc_NJP_2017_Massimo,elstc_2017_PRB_Ruzzene,elstc_2018_NM_LZY,elstc_PRB_2019_Ruzzene,elstc_PRAppl_2019_LMH,adma.202006521,elstc_JMPS_2020_HuangGL,Gigahertz_topological_valley,Wang2022_Extended_topological} waves. Regarding elastic wave, the valley-dependent phenomena, e.g., the valley edge states~\cite{elstc_2018_NM_LZY,elstc_2017_PRB_Ruzzene} and topological transportation~\cite{elstc_PRAppl_2019_LMH,elstc_PRB_2019_Ruzzene} are demonstrated experimentally. The topological phase transition of these lattice could be characterized by the valley Chern number~\cite{ValleyChernNmbr,elstc_JMPS_2020_HuangGL} and carried out by the honeycomb lattice with broken symmetry~\cite{Dirac_2013_PRB_Jos,PhysRevApplied.9.014001}. Due to topological protection, these elastic topological edge states (ETES) based on valley DOF are immune to sharp corners and structural defects~\cite{elstc_NJP_2017_Massimo,elstc_2018_NM_LZY}.

The elastic valley pseudospin-up (or -down) states of ETES are mathematically described by the effective Dirac Hamiltonian~\cite{elstc_2018_NM_LZY}. As such, the elastic pseudospin has been underlined with the help of chiral vibration~\cite{elstc_2018_NM_LZY} and the energy flux ~\cite{elstc_JMPS_2020_HuangGL,acstc_PRB_2017_LZY}, which deepens the understanding about the relationship between chiral vibration modes and pseudospin~\cite{PhysRevLett.119.255901,PMID:31549065}. However, pseudospin fails to represent the detailed properties of a finite inhomogeneous area that contains rich information. At given position, a local real physical quantity with measurable effects is desired to help people to identify or manipulate the ETES. 

Recently, research in optical~\cite{OptSpin_2014_NC_Bliokh,QSHEinLight,Leuchs-From_AM_To_PhotWh-NatPhot-2015,Bliokh-SO_light-NatPhot-2015}, acoustic~\cite{10.1093/nsr/nwz059,AcstcSpin_2019_PRB_Bliokh,AcstcSpin_2019_PRL_Bliokh,long2020realization,10.1093/nsr/nwaa040} and elastic~\cite{Long9951,Lanoy30186,yuan2021observation} systems show that the real physical quantity, spin angular momentum (spin for short) in classic waves plays an important role in the topological and unidirectional transportation. For example, the elastic spin demonstrates the local chirality locking mechanism of Rayleigh wave in conventional semi-infinite medium~\cite{Long9951, yuan2021observation}. The elastic spin is different from pseudospin~\cite{PhysRevLett.119.255901,PMID:31549065}, the latter of which is not spin angular momentum, but represents a $SU(2)$ spin-like mathematical structure in various systems. Pseudospin does not have the unit of angular momentum, but is a dimensionless quantity. In contrast, the elastic spin is a real physical quantity as angular momentum of the $SO(3)$ structure~\cite{Long9951,RJsingle}, which dose not rely on abstract analogue of $SU(2)$ mathematical structure.

Therefore, on one hand, although it is well accepted that the pseudospin-momentum locking is crucial for ETES, how the real local elastic spin locked with the directionality at different locations is not yet uncovered in topological elastic metamaterials. On the other hand, although the elastic spin could unveil more local physics in 3D complex structures, the application of elastic spin to study topological metamaterials is yet absent so far, let alone the valley elastic metamaterials. 

Here, we construct topological phononic crystal (PC) plates based on valley DOF, and create ETES in between PC plates with opposite valley Hall phases. Particularly, we introduce the elastic spin into the valley PC plates and show the local elastic spin-momentum locking mechanism in valley metamaterials. We observe the unidirectional wave propagation according to both the spin distribution of edge states and position of meta-sources with sub scale in PC plate. The direction of spin density inside the edge states can give a clear guidance for selectively chiral manipulation. Our results advance the knowledge of local features of topological meta-materials, build a bridge between topological meta-materials and intrinsic elastic spin, and provide a flexible tool to investigate spin related phenomena in integrative on-chip devices~\cite{cha2018experimental,elstc_2018_NM_LZY,adma.202006521}.

\begin{figure}[!htb]
  \centering
  \includegraphics[width=\linewidth]{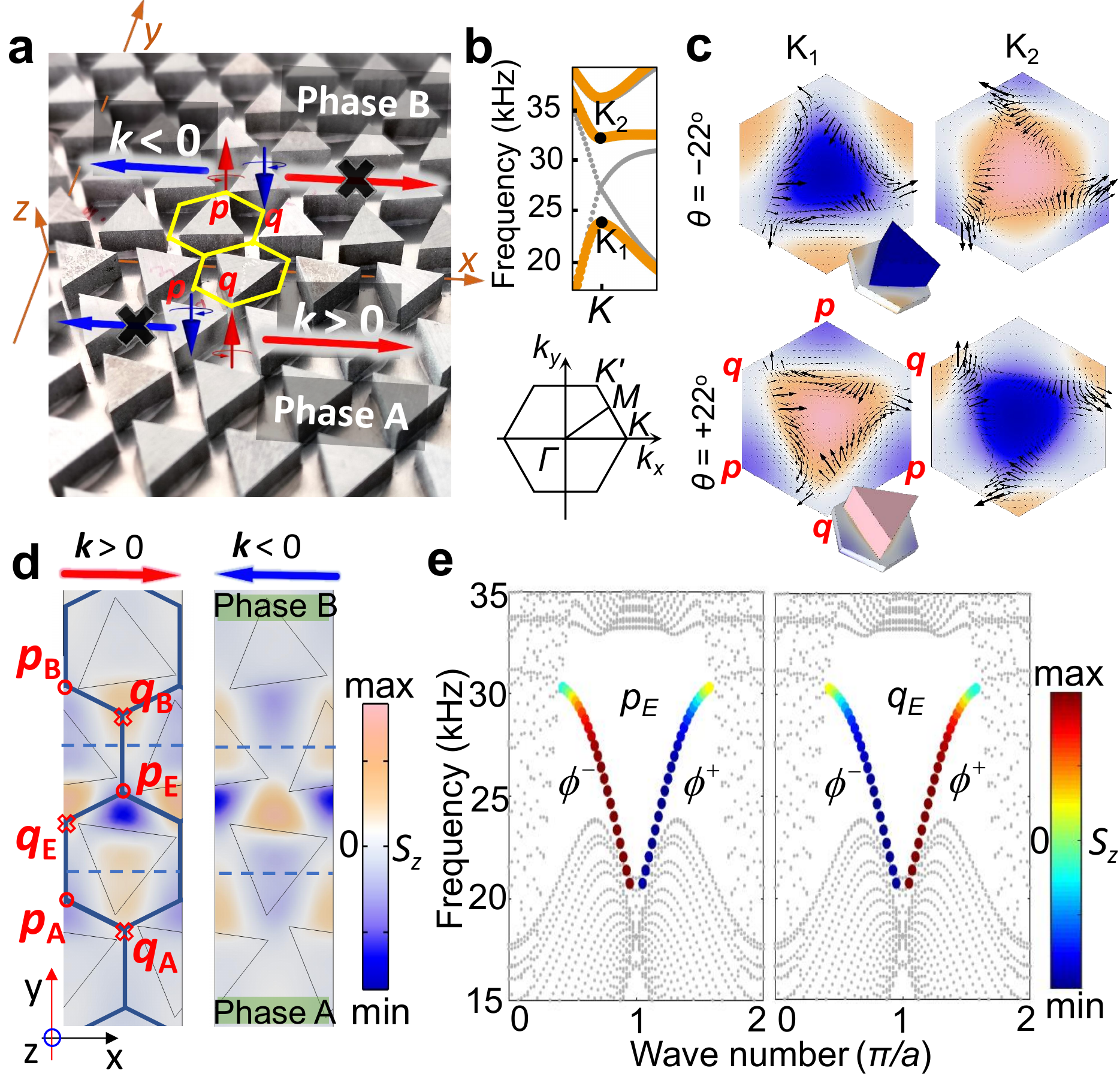}\vspace{-3mm}
  \caption{\textbf{Valley PC plate interface and scheme of elastic spin-valley locking mechanism.} \textbf{a} Photograph of experimental valley PC plate and scheme of elastic spin-valley locking mechanism. As an example, opposite spin sources at $p$ points can generate the opposite directional ETES, but opposite spin source at $p$ and $q$ point can generate the same directional ETES. \textbf{b} Band structures of PC plate when $\theta$ = $0^{\circ}$ (grey dots) and $\theta$ = $\pm22^{\circ}$ (orange dots). The local extreme values of first and second bands are labeled as $K_1$ and $K_2$, respectively. Bottom panel shows the first Brillouin zone of unit cell. \textbf{c} In-plane mechanical energy flux (black arrows) and  elastic spin density $S_z$ (twilight color table) at valleys $K_1$ (bottom) and $K_2$ (top) on top surface of substrate, respectively, when $\theta$ = $+22^{\circ}$ and $\theta$ = $-22^{\circ}$. (Insets) The $S_z$ distribution in unit cell. \textbf{d} Elastic spin density $S_z$ distribution on states of ETES with positive ($\pmb{k}>0$) and negative ($\pmb{k}<0$) group velocity at 25.8 kHz on substrate aluminum plate. The high-symmetry point around interface $\phi$ are labeled as $q_A$, $p_A$, $q_E$, $p_E$, $q_B$, $p_B$ in order. The subscripts $A$, $B$ or $E$ present the points in phase $A$, $B$ or along interface edge. \textbf{e} Dispersion of supercell containing interface $\phi$ with phase B ($\theta$ = $-22^{\circ}$) up and phase A ($\theta$ = $+22^{\circ}$) low. The color scale indicates the spin density obtained at $p_E$ and $q_E$ points, respectively.ETES branches $\phi^+$ and $\phi^-$ feature the positive and negative group velocity, respectively.
\label{fig1}
}
\end{figure}

Figure \textbf{\ref{fig1}a} shows the photograph of a valley PC plate fabricated by erecting triangular aluminum pillars in hexagonal lattice on an aluminum plate using epoxy adhesiv(see also Supplementary Fig.~\textbf{1a}~\footnote{See Supplemental Material at (URL)
about design of chiral elastic spin meta-source, sample fabrication, experimental measurement, numerical simulations, data processing and mass-spring model, which includes Refs.~\cite{1975Wave,yang2021abnormal,PhysRevB.98.094302,doi:10.1073/pnas.1507413112}\label{fn}}). The topological properties of PC plate with phase A and B are opposite with each other due to the converse rotation of pillars (Figs. \textbf{\ref{fig1}b} and \textbf{\ref{fig1}c}). Then, two ETES branches with positive and negative slopes occur in between the bulk band gap(Fig. \textbf{\ref{fig1}e})~\cite{elstc_2018_NM_LZY}.The primary interest here is unveiling the hidden spin momentum locking mechanism which predicts multiple selective excitation of ETES, as shown in Fig. \textbf{\ref{fig1}a}. For example, the spin-up source at $p$ point or spin-down source at $q$ point can generate the leftward ($\pmb{k}<0$) wave; while the reversed condition leads to the rightward ($\pmb{k}>0$) wave.

Figure \textbf{\ref{fig1}d} shows the spin density $S_z$ of ETES at 25.8 kHz on the substrate aluminum plate, where the elastic spin density is governed by ~\cite{Long9951, RJsingle} 
\begin{equation}\label{eq.szgeneral}
    \begin{aligned}
    \pmb{S}=\frac{\rho\omega}{2}\text{Im}(\pmb{u}^*\times\pmb{u}),
   \end{aligned}
\end{equation}
while the component along $z$ axis is then calculated by $S_z$=$\frac{\rho\omega}{2}\text{Im}(u_x^*u_y-u_xu_y^*)$. $\pmb{u}$ is the displacement vector. The time term is $\exp(-i\omega t)$ for monochromatic wave. The elastic spin is related to the spin angular momentum of media parcel with chirally polarized displacement vector. Elastic spin depends only on the displacement vector at selected location, which is different from the orbital angular momentum that relates also to the choice of coordinate origin~\cite{Long9951}. In this work, the clockwise (resp. anticlockwise) polarized displacement vector in $x$-$y$ plane yields the negative (resp. positive) spin density for $S_z$, respectively.

From Fig. \textbf{\ref{fig1}d}, the $p_A$, $p_B$ and $p_E$ points all have negative $S_z$ on the state of ETES with positive group velocity ($\pmb{k}>$ 0, $\phi^+$), but the $q_A$, $q_B$ and $q_E$ all have positive $S_z$. The $S_z$ shows reversed distribution on state of ETES with negative group velocity ($\pmb{k}<$ 0, $\phi^-$).
For comparison, Supplementary Fig. \textbf{3}~\footnotemark[1] shows the in-plane energy flux arrows near the interface that do not maintain distribution as clear as the elastic spin density. Note that we only show the spin density and distribution of energy flux on substrate, considering the spin meta-source is erected on the substrate.

To show the theoretical origin of $S_z$ distribution, we return to the basic topological properties of valley PC plates and ETES branches. Figure \textbf{\ref{fig1}b} shows the band structures of valley PC plate based on the milestone work~\cite{elstc_2018_NM_LZY}. When $\theta$ = $0^{\circ}$, two degenerate states occur at K and they are separated by the rotation of pillars, as shown in Fig. \textbf{\ref{fig1}b}. The positive (resp. negative) distribution of spin density $S_z$ near $p$ and $q$ points is similar to the anti-clockwise (resp. clockwise) polarization of $p$ and $q$ sites in the mass-spring model ~\footnotemark[1]. This is because the elastic spin~\cite{Long9951} is related to the mechanical angular momentum of media parcel~\cite{PhysRevLett.115.115502}. 
Supposing the medium parcel at $p$ and $q$ point as the mass and the rest of medium as the elastic springs, the spin of eigen states at $p$ and $q$ point are related to the circular motion of the eigenstates in mass-spring models~\cite{PhysRevLett.119.255901,PMID:31549065,HGK2019JMPS} that 
can be described by the effective Dirac Hamiltonian~\footnotemark[1].

When the $C_{3v}$ symmetry is broken by the rotation of triangle pillars, the eigen frequencies of two states at K are shifted to open a band gap for flexural plate wave. As shown in Fig. \textbf{\ref{fig1}c}, the frequency of eigen state with negative spin density at $p$ point decreases when $\theta$ = $22^{\circ}$ but increases when $\theta$ = $-22^{\circ}$. Regarding both $\theta$ values, the spin inside triangle pillars is reversed between $K_1$ and $K_2$. For example, when $\theta$ = $22^{\circ}$, $S_z$ in triangular pillar at $K_1$ is purely positive, indicating the anti-clockwise polarization of in-plane displacement field anywhere inside the pillar and therefore the anti-clockwise motion of single pillar. This is well consistent with the reported anti-clockwise motion of pillar of valley chiral state~\cite{elstc_2018_NM_LZY}. Furthermore and qualitatively, the spin direction (up or down) of $S_z$ at $p$ or $q$ point agrees with the in-plane energy flux rotation (anti-clockwise or clockwise) therein. This lies in the phase difference of $\pm2\pi$, e.g., the $u_z$ around $p$ or $q$ points have experienced $\pm2\pi$ phase difference~\cite{acstc_PRL_2016_LZY}. This applies also to meta-source where the input signal on PZT $S1$-$S4$ experiences $\pm2\pi$ phase difference ~\footnotemark[1].

Now let us combine PC plates with opposite rotational angle, i.e phase B up and A down. Figure \textbf{\ref{fig1}e} shows the projected dispersion curves of supercell of $\phi$, magnifying two ETES branches with forward $(\phi^+)$ or backward $(\phi^-)$ group velocity, respectively. The distribution of $S_z$ along both ETES is already shown in Fig. \textbf{\ref{fig1}d}, while Fig.~\textbf{\ref{fig1}e} further summarizes the $S_z$ value along both ETES branches, e.g. at $p_E$ (left panel, also for $p_A$ and $p_B$) and $q_E$ (right panel, also for $q_A$ and $q_B$), respectively. Figure \textbf{\ref{fig1}e} yields a novel and important deduction: the $S_{pz}<0$ or $S_{qz}>0$ located at $p$ or $q$, respectively, is tightly bonding to the $\phi^+$ ETES due to spin-momentum locking, while the $S_{pz}>0$ or $S_{qz}<0$ located at $p$ or $q$ allows for $\phi^-$ ETES, in the frequency range of ETES with nearly linear slope.

The elastic spin at $p$ and $q$ points near the interface can be qualitatively understood by analytically solving the simplified mass-spring model~\footnotemark[1], as:
\begin{equation}
    S_{pz} =-\frac{\rho \omega \tau}{4} e^{-2\lambda \left| y \right|}, \quad\quad
    S_{qz} =\frac{\rho \omega \tau}{4} e^{-2\lambda \left| y \right|}.
\end{equation}
$S_{pz}$ and $S_{qz}$ are elastic spin density derived from the displacement vectors of $p$ and $q$ sites. $\lambda$ denotes the decaying rate of the ETES apart from the interface. $\tau$ = +1 (-1) is associated with $K$ ($K^{'}$) valley or $\pmb{k}>0$ ($\pmb{k}<0$). Obviously, the sign of elastic spin density depends on propagation direction of ETES, and $S_{pz}$ and $S_{qz}$ are opposite with each other, magnifying well the calculated results in Figs.~\textbf{\ref{fig1}d} and~\textbf{\ref{fig1}e}.

\begin{figure}[!htb]
  \centering
  \includegraphics[width=\linewidth]{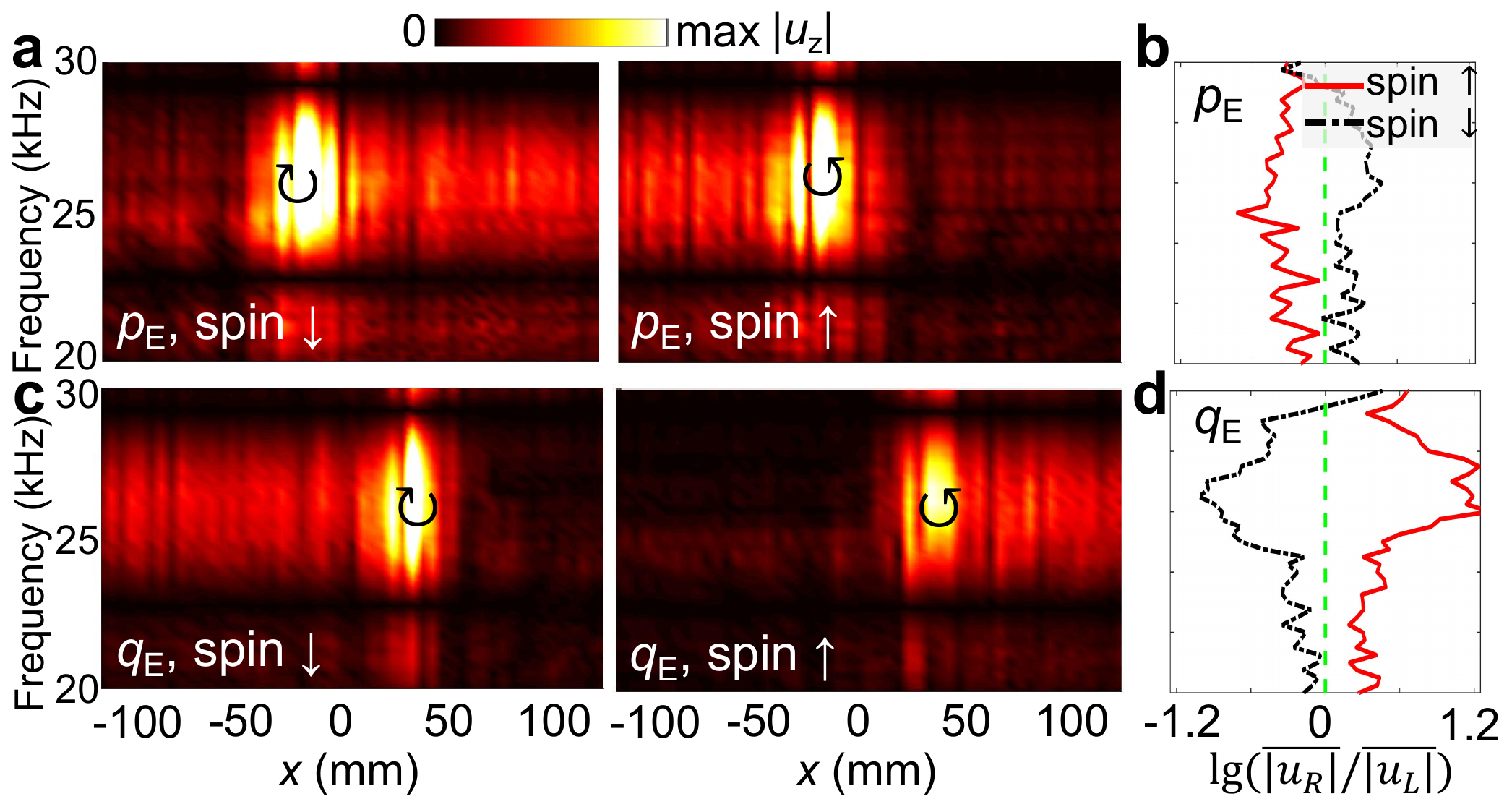}
  \caption{\textbf{Experimental observation of unidirectional wave transportation}. \textbf{a} The mapping of normalized $|u_z|$ measured along interface $\phi$ from $x$ = -125 mm to 125 mm every 5 mm  with elastic spin meta-source centered at $p_E$, where $x$ = 0 is at the middle position along interface $\phi$. \textbf{b} The difference ratio of $|u_z|$ between right and left side of meta-source, defined as lg$(\overline{|u_{R}|}/\overline{|u_{L}|})$, where $\overline{|u_{R}|}$ (resp. $\overline{|u_{L}|}$) is the averaged amplitude of $|u_z|$ measured at the right (resp. left) side of source, when meta-source is centered at $p_E$. The red solid and black dashed lines present the result with spin-up (spin $\uparrow$) and spin-down (spin $\downarrow$) sources, respectively. \textbf{c,d} The mapping of normalized $|u_z|$ (\textbf{c}) and difference ratio (\textbf{d}) measured along interface $\phi$ with meta-source locating at $q_E$.}
\label{fig3}
\end{figure}

For experimental measurement, we put up elastic spin meta-source centered at $p_A$, $q_A$, $p_E$, or $q_E$ with both spin-up and spin-down settings. The meta-source is made of four small lead zirconate titanate piezoelectric ceramics (PZT) disks with radii $r=2.5$ mm arranged in square lattice~\cite{Ye2017observation} (see Supplementary Figs. \textbf{1c} and \textbf{1d}~\footnotemark[1]). Although the assembly of the meta-source occupies area about 11 mm$\times$11 mm, the effective area of meta-source is the center position (about 2.5 mm$\times$2.5 mm) with highly confined $S_z$ of the same sign. Figure \textbf{\ref{fig3}a} shows the $|u_z|$ measured along interface $\phi$ with meta-source central at $p_E$. Notice that for each measurement position, the amplitude of $u_z$ is retrieved from the time resolved profile by using Fourier transformation (FFT) of the recorded wave package at the target frequencies~\cite{yuan2021observation}. In Fig. \textbf{\ref{fig3}a}, the asymmetric rightward and leftward wave propagation are successfully generated by spin-down and spin-up source, respectively. This asymmetric propagation is observed in broad frequency range, as shown by the ratio lg$(\overline{|u_{R}|}/\overline{|u_{L}|})$ in Fig. \textbf{\ref{fig3}b}: the positive (resp. negative) value indicates the rightward (resp. leftward) wave propagation generated by the spin down (resp. up) source. These experimental results agree with and must be ascribed to the $S_z$ distribution in Fig. \textbf{\ref{fig1}e}: at $p_E$, the $S_z$ is negative for $\phi^+$ but positive for $\phi^-$. The spin-down (resp. spin-up) source central at $p_E$ with $S_z<0$ (resp. $S_z>0$) generates the $\phi^+$ (resp. $\phi^-$). Hence, the unidirectional transport is determined by both the source spin $S_z$ and the local elastic spin density distribution around interface $\phi$, which demonstrates strongly the elastic spin-momentum locking in valley topological phononic materials.

To further shed lights on this spin-momentum locking, we show in Figs. \textbf{\ref{fig3}c} and \textbf{\ref{fig3}d} the mapping of $|u_z|$ and the difference lg$(\overline{|u_{R}|}/\overline{|u_{L}|})$ along interface $\phi$, with the meta-source central at $q_E$. In this case, the leftward and rightward wave propagation are realized by using spin-down and up source, respectively, which is opposite to the situation shown in Fig. \textbf{\ref{fig3}a}. Notably, the amplitude of asymmetric elastic wave between right and left side can 
reach a difference of $15.8$ times (lg$(\overline{|u_{R}|}/\overline{|u_{L}|})=1.2$). Supplementary Figs. \textbf{6a}-\textbf{6d}~\footnotemark[1] detail the experimental results when the source is central at $p_A$ (resp. $q_A$), and the spin source generates asymmetric transportation in a way similar to those results by putting the source central at $p_A$ (resp. $q_A$). The observation of broadband unidirectional ETES agrees well with the $S_z$ distribution in Figs. \textbf{\ref{fig1}b} and \textbf{\ref{fig1}e}.

It is noteworthy that Fig. \textbf{\ref{fig3}} excludes frequencies up 30 kHz, considering that ETES branches are nearly flat with small spin density at $p$ or $q$, together with the possible perturbing by some bulk modes, as shown in Fig. \textbf{\ref{fig1}e}. The calculated ratio lg$(\overline{|u_{R}|}/\overline{|u_{L}|})$ (see Supplementary Fig.~\textbf{8} ~\footnotemark[1]) shows similar results above 30 kHz. In addition, the relatively weak asymmetric propagation at lower frequencies ($<\sim$24 kHz) may come from the experimental factors, e.g. installation errors of meta-source, and the possible influences from bulk modes. Besides, we must note that when using even smaller PZT disk (with radii $r=$1.5 mm) to construct meta-source, the installation process introduces even strong imperfection and downloads the ratio {lg}$(\overline{|u_{R}|}/\overline{|u_{L}|})$, e.g., at $p_A$ position. Yet, the broadband asymmetric wave transportation is still observable (see Supplementary~\footnotemark[1]).

\begin{figure}[!htb]
  \centering
  \includegraphics[width=\linewidth]{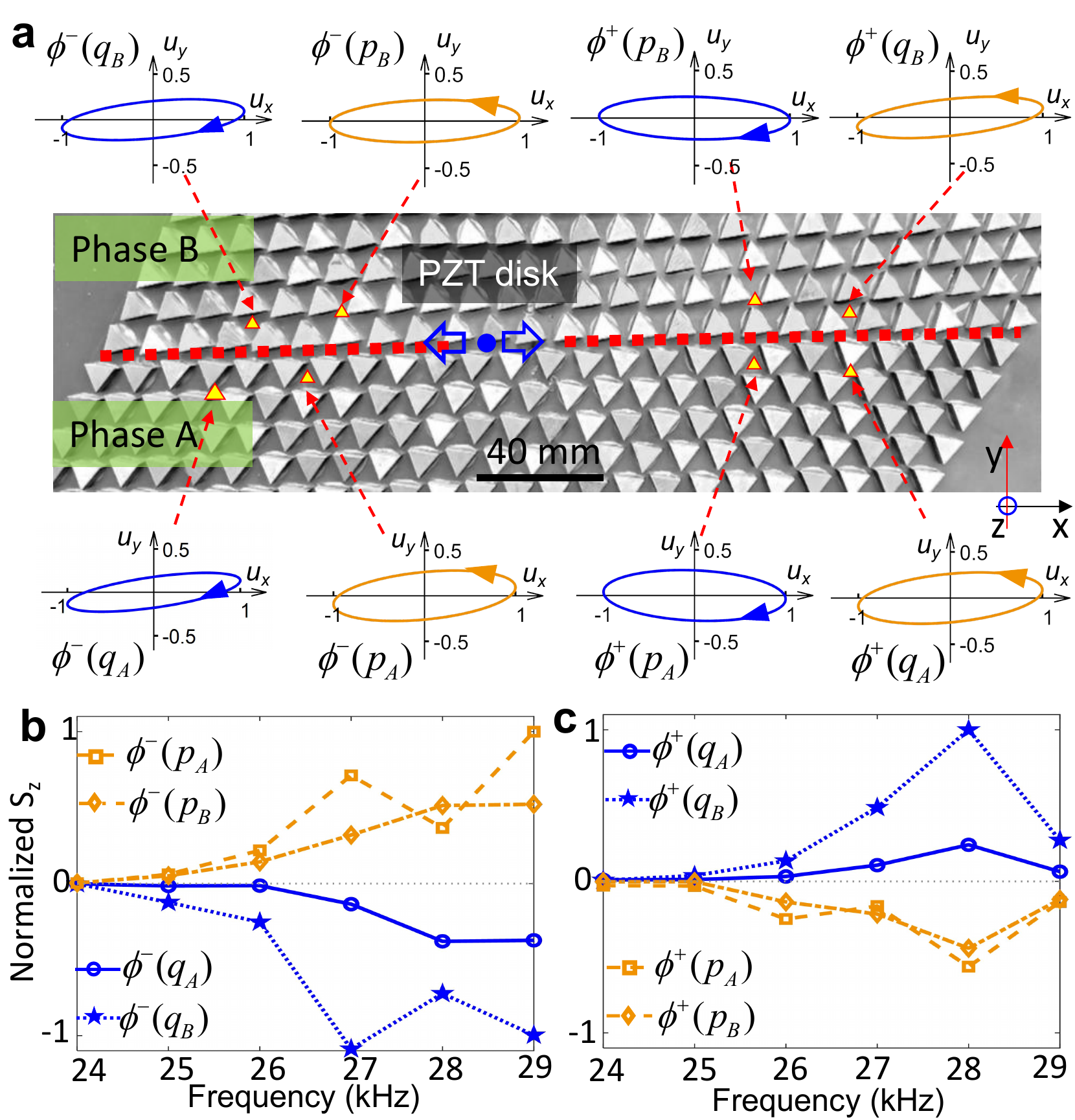}
  \caption{\textbf{Measured displacement $\pmb{u}$=$(u_x,u_y)$ polarization  and} elastic $S_z$ at representative $p$ and $q$ points. \textbf{a} Zooming in of photograph of experimental sample, and the measured displacement polarization at each $p$ and $q$ point in phase A and B, at $28$ kHz. A single PZT disk is set up around interface. \textbf{b,c} Measured $S_z$ at the left (\textbf{b}) and right (\textbf{c}) side of PZT source, normalized to the maximum in each panel. Experiments show that in both phases A and B, the elastic spins at locations $p$ and $q$ feature opposite spins, which further reverse the sign when the propagation direction of ETES is reversed. This observation clearly demonstrates the hidden elastic spin-momentum locking and elastic spin-valley locking.}
\label{fig4}
\end{figure}

Finally, we turn to the experimental characterization of $S_z$ at $p$ or $q$ points. Figure \textbf{\ref{fig4}a} shows the zooming in of experimental sample for interface $\phi$. Without losing generality, one PZT disk is chosen to generate both $\phi^+$ and $\phi^-$ instead of meta-source. The yellow triangles stand for the position of subwavelength triangular pillars at high-symmetry points ($p_A$, $q_A$, $p_B$, and $q_B$) but on the bottom surface of PC plate. By measuring the displacement that is perpendicular to the side surface of subwavelength triangular pillars (see Supplementary~\footnotemark[1]), we can obtain the $u_x$ and $u_y$ at the selected points. After making FFT of $u_x$ and $u_y$, we can extract the amplitude and phase information of FFT components of $u_x$ and $u_y$ at the given frequency, e.g. 28 kHz. Then, we draw the displacement polarization by using the retrieved amplitude and phase value of $u_x$ and $u_y$ at the given frequency, and obtain the elastic spin density~\cite{yuan2021observation}.
Figure \ref{fig4}a shows the experimentally obtained displacement polarization at both $p$ and $q$ points. At the left side of wave source, the $\phi^-(p_A)$ and $\phi^-(p_B)$ feature anticlockwise polarized $\pmb{u}$, resulting in positive $S_z$. While the $\phi^-(q_A)$ and $\phi^-(q_B)$ feature clockwise polarized $\pmb{u}$ with the negative $S_z$. Regarding the right side of wave source, the situation is reversed: both $p$ sites, i.e. $\phi^+(p_A)$ and $\phi^+(p_B)$, show clockwise polarized $\pmb{u}$ with negative $S_z$, while both $q$ sites, i.e. $\phi^+(q_A)$ and $\phi^+(q_B)$, demonstrate the anticlockwise polarized $\pmb{u}$ with positive $S_z$.

Figure \textbf{\ref{fig4}b} summarizes the measured $S_z$ at points $p$ and $q$ at the left side of source. In short, the $S_z$ features positive value on $p$ points but negative value on $q$ points within broad frequency range. The situation at points $p$ or $q$ at right side of source, as presented in Fig. \textbf{\ref{fig4}c}, shows the reverse scenario: the $S_z$ is negative on $p$ points but positive on $q$ points. These experimental results agree well with the numerical ones shown in Supplementary. In Figs. \textbf{\ref{fig4}b} and \textbf{\ref{fig4}c}, the normalized $S_z$ features smaller value than the numerical counterparts (Supplementary Fig.~\textbf{8}~\footnotemark[1] ) at lower frequency, due to the imperfect installation of source, the possible influence of bulk edge modes and some other factors in experiment. Nevertheless, both experimental and numerical results confirm that the unidirectional wave transportation is spin-momentum locking between the spin of meta-source and the spin density distribution on PC plate at discrete points.

In conclusion, we have successfully observed the elastic spin density at selected local points in valley topological materials and the hidden elastic spin-valley locking resulted unidirectional transport. The selective chiral coupling to the propagating direction is mostly determined by the local spin-momentum locking according to the elastic spin of source and local spin density distribution in valley topological PC plate. In the future, the real measurable physical effects of elastic spin angular momentum, such as the spin-torque correspondence~\cite{AcstcSpin_2019_PRL_Bliokh,10.1093/nsr/nwz059,RealSpin_JAP}, could largely expand the application scenarios of topological valley phononic crystals. The local elastic valley spin as well as the hidden elastic spin-valley locking advances the knowledge about the conventional features in topological meta-materials and phononics~\cite{PhysRevLett.105.225901,phononics2012}, which can be extended to other topological states and offers flexible tool to build up elastic devices like elastic (or phonon) logic devices~\cite{boechler2011bifurcation,PhysRevLett.119.255901,phonon_diode_NPht_2022_ZhangLiFa},  topological nonreciprocal transmitter~\cite{PhysRevB.101.085108,nassar2020nonreciprocity}, unidirectional information transportation~\cite{Ye2017observation,acstc_PRB_2017_LZY,yuan2021observation,UniPhonon_PRL_2020_YuTao}, intriguing transport in novel valley system~\cite{PhysRevLett.128.216403,acstc_PRL_2018_LZY}, and etc.

This work is supported by National Natural Science Foundation of China (Nos. 11935010, 12172256 and 11872282), Shanghai Science and Technology Committee (No. 20ZR1462700), and the Opening Project of Shanghai Key Laboratory of Special Artificial Microstructure Materials and Technology.

\bibliography{ref}

\newpage

\title{Elastic Valley Spin Controlled Chiral Coupling in Topological Valley Phononic Crystals-Supplementary}

\author{Jinfeng Zhao}
\thanks{
These authors contributed equally: Jinfeng Zhao, Chenwen Yang, Weitao Yuan}
\affiliation{School of Aerospace Engineering and Applied Mechanics, Tongji University, 100 Zhangwu Road, Shanghai 200092, China}

\author{Chenwen Yang}
\thanks{
These authors contributed equally: Jinfeng Zhao, Chenwen Yang, Weitao Yuan}
\affiliation{Center for Phononics and Thermal Energy Science, China-EU Joint Lab on Nanophononics, Shanghai Key Laboratory of Special Artificial Microstructure Materials and Technology, School of Physics Science and Engineering, Tongji University, Shanghai 200092, China}

\author{Weitao Yuan}
\thanks{
These authors contributed equally: Jinfeng Zhao, Chenwen Yang, Weitao Yuan}
\affiliation{School of Aerospace Engineering and Applied Mechanics, Tongji University, 100 Zhangwu Road, Shanghai 200092, China}

\author{Danmei Zhang}
\affiliation{Center for Phononics and Thermal Energy Science, China-EU Joint Lab on Nanophononics, Shanghai Key Laboratory of Special Artificial Microstructure Materials and Technology, School of Physics Science and Engineering, Tongji University, Shanghai 200092, China}

\author{Yang Long}
\affiliation{Center for Phononics and Thermal Energy Science, China-EU Joint Lab on Nanophononics, Shanghai Key Laboratory of Special Artificial Microstructure Materials and Technology, School of Physics Science and Engineering, Tongji University, Shanghai 200092, China}

\author{Yongdong Pan}
\affiliation{School of Aerospace Engineering and Applied Mechanics, Tongji University, 100 Zhangwu Road, Shanghai 200092, China}

\author{Hong Chen}
\affiliation{Center for Phononics and Thermal Energy Science, China-EU Joint Lab on Nanophononics, Shanghai Key Laboratory of Special Artificial Microstructure Materials and Technology, School of Physics Science and Engineering, Tongji University, Shanghai 200092, China}

\author{Zheng Zhong}
\email{
Corresponding address: zhongk@tongji.edu.cn}
\affiliation{School of Aerospace Engineering and Applied Mechanics, Tongji University, 100 Zhangwu Road, Shanghai 200092, China}

\author{Jie Ren}
\email{
Corresponding address: Xonics@tongji.edu.cn}
\affiliation{Center for Phononics and Thermal Energy Science, China-EU Joint Lab on Nanophononics, Shanghai Key Laboratory of Special Artificial Microstructure Materials and Technology, School of Physics Science and Engineering, Tongji University, Shanghai 200092, China}

\maketitle

\tableofcontents

\section{Design of chiral elastic spin meta-source.} Our chiral elastic spin meta-sources are composed of four subwavelength PZT disks that are fixed to the thin aluminum plate by epoxy adhesive in square. We first evaluate the chiral elastic spin meta-source in theory. Consider a thin solid plate in $x$-$y$ plane, the plate wave equation is then governed by~\cite{1975Wave}
\begin{equation}\label{eq.wave_Cartesian}
    \begin{aligned}
    \nabla^4u_z(x,y,t)- \omega^2\frac{\rho h}{D}u_z(x,y,t)=f(x,y,t)
   \end{aligned}
\end{equation}
where $\nabla^4$ is the biharmonic operator, $u_z(x,y,t)$ is the out-of-plane displacement along $z$ axis, $\omega$ is the angular frequency, $\rho$ is the mass density of plate, $h$ is the plate thickness, $D=\frac{Eh^3}{12(1-\mu^2)}$ is the plate bending stiffness, $E$ is the Young's modulus, $\mu$ is the shear modulus, and $f(x,y,t)$ is the applied force along $z$ axis.

For monochromatic wave, the out-of-plane displacement caused by the point source in thin plate can be assumed as
\begin{equation}\label{eq.uz}
    \begin{aligned}
    u_{zj}=u_{z0}e^{i(kr_j-\omega t)}e^{-\gamma r_j}e^{i\phi_j}
   \end{aligned}
\end{equation}
where $u_{z0}$ is the initial amplitude, $k$ is wave number of plate wave, $\gamma$ is a positive number indicating the decay of propagating wave away from the point source, $\phi_j$ is an additional phase, $r_j$ = $\sqrt{(x-x_j)^2-(y-y_j)^2}$ and ($x_j$,$y_j$) is the $j^{th}$ point source location. Notice that near-field evanescent wave of the point source is ignored in this equation. 

Typically, for a thin plate in $x$-$y$ plane, the stress components $\tau_{yz}$, $\tau_{zx}$, and $\sigma_{z}$ are much smaller than other components $\sigma_x$, $\sigma_y$ or $\tau_{xy}$. As a consequence, the strain components relating to $\tau_{yz}$ and $\tau_{zx}$ satisfy condition
\begin{equation}\label{eq.epsilon}
    \begin{aligned}
    \varepsilon_{yz}=\frac{1}{2}(\frac{\partial u_z}{\partial y}+\frac{\partial u_{y}}{\partial z})=0\\
    \varepsilon_{zx}=\frac{1}{2}(\frac{\partial u_x}{\partial z}+\frac{\partial u_z}{\partial x})=0
   \end{aligned}
\end{equation}
where $u_x$ and $u_y$ are the in-plane displacement along $x$ and $y$ axes, respectively. Considering $u_{zj}$ is only the function of $x$ and $y$, the in-plane displacement can be obtained as
\begin{equation}\label{eq.uxuy}
    \begin{aligned}
    u_{xj}=-z\frac{\partial u_{zj}}{\partial x}=-z(ik-\gamma)u_{z0}\frac{x-x_j}{r_j}e^{i(kr_j-\omega t)}e^{-\gamma r_j}e^{i\phi_j}\\
    u_{yj}=-z\frac{\partial u_{zj}}{\partial y}=-z(ik-\gamma)u_{z0}\frac{y-y_j}{r_j}e^{i(kr_j-\omega t)}e^{-\gamma r_j}e^{i\phi_j}.
   \end{aligned}
\end{equation}
Here, the $z=0$ is set at the middle plane of thin plate.

The elastic spin density in isotropic solids is governed by ~\cite{Long9951}
\begin{equation}\label{eq.szgeneral}
    \begin{aligned}
    \pmb{S}=\frac{\rho\omega}{2}Im(\pmb{u}^*\times\pmb{u}).
   \end{aligned}
\end{equation}
The component of $\pmb{S}$ along the $z$ axis $S_z$ is then calculated by $S_z$=$\frac{\rho\omega}{2}Im(u_x^*u_y-u_xu_y^*)$. When elaborating the meta-source by using multiple point sources, e.g., two point sources, $S_z$ can then be written as 
\begin{equation}\label{eq.sz}
    \begin{aligned}
    S_z=\frac{\rho\omega}{2}Im[(u_{xj}+u_{xl})^*(u_{yj}+u_{yl})\\-(u_{xj}+u_{xl})(u_{yj}+u_{yl})^*]=\sum_{j,l=1,2}S_{z(jl)}
   \end{aligned}
\end{equation}
where
\begin{equation}\label{eq.szjl}
    \begin{aligned}
    S_{z(jl)}=\frac{\rho\omega}{2}Im[(u_{xj})^*(u_{yl})-(u_{xj})(u_{yl})^*]\\
    =\frac{\rho\omega}{2}(zu_{z0})^2(k^2+\gamma^2)\frac{(x-x_j)(y-y_l)}{r_jr_l}e^{-\gamma(r_j+r_l)}\\\times Im[e^{-ik(r_j-r_l)}e^{-i(\phi_j-\phi_l)}-e^{ik(r_j-r_l)}e^{i(\phi_j-\phi_l)}]
   \end{aligned}
\end{equation}
with * as the conjugate operator. When the meta-source is built by four point sources, the $S_z$ can be derived as $S_z = \sum_{j,l=1,2,3,4}S_{z(jl)}$.

To give illustration, in Supplementary Figures \textbf{\ref{figs3}} and \textbf{\ref{figs4}}, we show the theoretical results of meta-source consisting four point sources located as ($x$,$y$) = (-3 mm, 3 mm), (3 mm, 3 mm), (3 mm, -3 mm), (-3 mm, -3 mm) with additional phase $\phi$ = $0$, $\pm\pi/2$, $\pm\pi$, $\pm3\pi/2$, in accordance to the experimental configuration. $\gamma$ is set as 0 for simplicity. Obviously, $S_z$ presents good agreement between the theoretical, numerical and experimental results, and so is the $u_z$ distribution. It is noteworthy that $S_z(jl)$ includes $z^2$ so that the value of $S_z$ present the same sign along the thickness direction, with the maxima on both the top and bottom surfaces at $z=-h/2$ and $z=h/2$, respectively. 

\section{Sample Fabrication.} The phononic crystal (PC) plate is fabricated by erecting triangular pillars on substrate plate using epoxy adhesive. The triangular pillars and substrate plate are made of 6061-T6 aluminum. We first perforate triangular pillars from a 5 mm thick aluminum plate with geometric side length $s$ = 10 mm by laser cutting technique. Then we manually fix an array of 25 $\times$ 22 triangular pillars in hexagonal lattice to a 800 mm $\times$ 700 mm rhombus aluminum plate with epoxy adhesive. The rotation angle of pillars is $\theta=22^{\circ}$ and $-22^{\circ}$, respectively. Meanwhile, the PC occupies a rhombus region to reduce the influence of wave reflection between PC plate and pure aluminum plate. After 24 hours at room temperature, the strength of the epoxy adhesive reaches their maximum, and these samples can be used for experiment. 

\section{Experimental measurement.} To generate ultrasonic wave pulses, the subwavelength lead zirconate titanate piezoelectric ceramic (PZT) disks are used to excite point-like source and chiral elastic spin meta-source. The PZT disks are fixed to substrate plate by using epoxy adhesive. In both cases, the five-cycled tone burst electrical signals with central frequency $f_c$ are issued from the function generator (RIGOL DG1032z) and then amplified by the power amplifier (Aigtek ATA-2022H). The every $\pm\pi/2$ phase difference against $f_c$ in chiral meta-source is tuned by using function generator (RIGOL DG1032z). To measure elastic wave propagation, the laser Droppler vibrometer (LDV, Polytec vibrometer OFV 2570) is used to
record the out-of-plane displacement $u_z$ at any position on substrate surface, and the out-of-plane displacement perpendicular to each face of small triangular pillars for elastic SAM measurement (Supplementary Fig. \textbf{\ref{figs3}}). Each measurement comes from the averaging
over 128 scans digitized by an oscilloscope (DPO 4102B) at sampling frequency of 250 MHz. This measurement allows for a good signal-to-noise (S/N) ratio. 

\section{Numerical simulations.} The numerical simulations are performed by finite element method using commercial software COMSOL Multiphysics. For the physical parameters of 6061-T6 aluminum, we chose the mass density $\rho_{Al}=2700 kg/m^3$, Young's modulus $E_{Al}=67.7$ GPa, and Poisson's ratio $\nu_{Al}=0.35$. For the epoxy adhesive, we set the $\rho_{epoxy}=1.2 kg/m^3$, $E_{epoxy}=5.1$ GPa, and $\nu_{epoxy}=0.38$. The geometric parameters include $a$ = 12 mm, $e$ = 1.5 mm, $h_g$ = 0.14 mm, $h_p$ = 5 mm, $s$ = 10 mm, $\theta$ = $\pm22^{\circ}$.

\textcolor{black}{\section{Data processing.} To get the displacement polarization and the maps of wave transportation, our experimental methods and detailed data analysis procedures are as what follows. In order to obtain the displacement polarization, we adopt an experimental method as shown in Supplementary Fig. \textbf{\ref{figs3}d}. We erect small sub-wavelength aluminum triangular pillars on the back surface of substrate, central at selected $p$ or $q$ point in phase A (B) or on interface. These small size pillars feature the side length 2.9 mm and height 3 mm that is much smaller than the triangular scatters for building PC plate. Then we use LDV to record the displacement that is perpendicular to the side surface of sub-wavelength small triangular pillars, as shown by the red arrows in Supplementary Fig. \textbf{\ref{figs3}d}. We label the displacement on $+x$ side as $u_1$ with angle $\theta_1$, while the displacement on $-x$ side as $u_2$ with angle $\theta_2$, so that $u_x$ and $u_y$ of the effective point can be obtained as $u_x = u_1cos(\theta_1)-u_2cos(\theta_2)$, and $u_y=u_1sin(\theta_1)+u_2sin(\theta_2)$, in the form of time evolution signal. Then by making the Fourier transformation (FFT) of $u_x$ and $u_y$ within the same time period including the same target wave package, we can extract the amplitude and phase information of FFT components of $u_x$ and $u_y$ at given frequency, e.g. 28 kHz. After that, we can draw the displacement polarization by using the retrieved amplitude and phase value of $u_x$ and $u_y$ at the given frequency with time term $e^{-i\omega t}$, and we obtain the elastic SAM density. To draw the maps of normalized $|u_z|$ along interface, we measure the out-of-plane displacement $u_z$ at a series of points along interface $\phi$ on substrate, from $x=-$125 mm to 125 mm every 5 mm, and get the time evolution signal at each point. Then we make the FFT of measured time evolution signal and extract the amplitude information of FFT components of $u_z$ in concerned frequency range. The assembly of all the extracted amplitude of $u_z$ gives rise to the maps of $u_z$ in Figs. \textbf{3a} and \textbf{3c}, and Supplementary Figs. \textbf{\ref{figs5}} and \textbf{\ref{figs6}}. }

\textcolor{black}{\section{The mass-spring model.}  The SAM density of elastic wave is related with the mechanical angular momentum of the medium parcel which has a circularly polarized displacement vector~\cite{Long9951}. 
We can understand the SAM properties of $p$ and $q$ points through the well-established mass-spring model~\cite{HGK2019JMPS}, where the mass-spring model is widely used to build and understand physical models~\cite{yang2021abnormal,elstc_NJP_2017_Massimo,Dirac_2013_PRB_Jos,PhysRevB.98.094302,elstc_2017_PRB_Ruzzene,doi:10.1073/pnas.1507413112}. Suppose the valley metamaterial in our work as a mass-spring system in Supplementary Fig. \textbf{\ref{figs10}}. We regard the medium parcel at $p$ and $q$ points as the masses and the rest medium as elastic springs.} 

\textcolor{black}{In Supplementary Fig. \textbf{\ref{figs10}}, the unit cell has a pair of mass $m_p$ and $m_q$ at $p$ and $q$ point, respectively, and the springs between them. The stiffness of spring is $g$, and the length of spring is $L_0$. Lattice vectors are $\bm{a_1}=(a_x,a_y )=(a/2,\sqrt{3}a/2)=(\sqrt{3}L_0/2,3L_0/2)$ and $\bm{a_2}=(a_x,-a_y )=(a/2,-\sqrt{3}a/2)=(\sqrt{3}L_0/2,-3L_0/2)$. Consider a Bloch wave of unit cell ($m$,$n$) in the form: $\bm{u}_{m, n}=\bm{u}e^{i\bm{k}\cdot (m\bm{a_1}+n\bm{a_2})}$, where $\bm{u}=[u_{px},u_{py},u_{qx},u_{qy}]$. 
The harmonic time term $e^{-i\omega t}$ is omitted here. By solving the equation of motion~\cite{HGK2019JMPS}, we can get the Hamiltonian of this model}
\textcolor{black}{
\begin{gather}
\omega^2 \bm{u}=H( {\bm{k}})\bm{u};\\
    H( {\bm{k}})= {\rm -\bm{M}}^{-1} {\rm \bm{K}}( {\bm{k}});\\
     {\rm \bm{M}}=\left[\begin{array}{cc}
         m_p  {\rm \bm{I}}&0  \\
         0 & m_q{\rm \bm{I}}
    \end{array}\right];
     {\rm \bm{K}}=\left[\begin{array}{cc}
          {\rm \bm{K}}_{pp} &  {\rm \bm{K}}_{pq}  \\
          {\rm \bm{K}}_{pq}^{\dag} & {\rm \bm{K}}_{qq}
    \end{array}\right];\\
       {\rm \bm{K}}_{pp}= {\rm \bm{K}}_{qq}=-\frac{3g}{2}\rm \bm{I};\\
     {\rm \bm{K}}_{pq}=g\left[\begin{array}{cc}
         \frac{3}{2}e^{-i k_y a_y} \cos(k_x a_x)  &
         -i \frac{\sqrt{3}}{2} e^{-i k_y a_y} \sin(k_x a_x)\\
         -i \frac{\sqrt{3}}{2} e^{i k_y a_y} \sin(k_x a_x) &1+\frac{1}{2} e^{i k_y a_y} \cos(k_x a_x)
    \end{array}\right].
\end{gather}
The subscript
$^\dag$ refers to the conjugate transpose, and $\rm \bm{I}$ is the unit matrix.
At Brillouin zone corner $\bm{K}$, the two eigen states of this Hamiltonian are
\begin{equation}
    \bm{u}_1(\bm{K})=\frac{1}{\sqrt{2}}(i,1,0,0)^{T};\bm{u}_2(\bm{K})=\frac{1}{\sqrt{2}}(0,0,-i,1)^{T},
\end{equation}
while at $\bm{K}^{'}$, the two eigen states can be written as
\begin{equation}
    \bm{u}_1(\bm{K}^{'})=\frac{1}{\sqrt{2}}(-i,1,0,0)^{T};\bm{u}_2(\bm{K}^{'})=\frac{1}{\sqrt{2}}(0,0,i,1)^{T}.
\end{equation}
}\textcolor{black}{These eigen states show a clear picture of the circularly polarized displacements of mass. For example, at $\bm{K}$, $\bm{u}_1(\bm{K})=\frac{1}{\sqrt{2}}(i,1,0,0)^T$ means $u_{px}/u_{py}=i=e^{i\pi/2}$, and the displacement vector at $p$ shows clockwise polarization but equates zero at $q$ point. Thus, in continuum valley metamaterial, we shall get the negative SAM density at $p$ point and the zero SAM density at $q$ point. According to above theoretical eigen states, we can also see that at $\bm{K}$, one state possess clockwise polarization at $p$ site ($\bm{u}_1(\bm{K})$) while the other state shows the anticlockwise polarization at $q$ site ($\bm{u}_2(\bm{K})$), explaining well the calculated negative SAM density at $p$ point ($\bm{K}_1$) and positive SAM density at $q$ point ($\bm{K}_2$) in continuum valley metamaterial.
}

\textcolor{black}{The effective Hamiltonian near the Dirac point of this system is expressed as
\begin{gather}
\Delta \bm{H} \psi=\Delta \omega \psi;\\
\Delta \bm{H}=m \bm{\sigma}_z+v(\tau\Delta k_x \bm{\sigma}_x-\Delta k_y \bm{\sigma}_y);\\
m=\frac{1}{2} \frac{m_q-m_p}{m_q+m_p} \omega_0; v=\frac{\omega_0 a}{4 \sqrt{3}}; \omega_0=\sqrt{\frac{3 g}{m_p+m_q}},
\end{gather}
where $\tau =+1(-1)$ relates to the $\bm{K}(\bm{K}^{'})$ valley, $\psi=(c_1,c_2)^{T}$ is eigenvector related to the linear combination of degenerated states at $\bm{K}$, i.e. $\bm{u}(\bm{K})=c_1 \bm{u}_1(\bm{K})+c_2 \bm{u}_2(\bm{K})$.}

\textcolor{black}{Consider an interface along $x$-axis at $y=0$ ($y<0,m<0;y>0,m>0$), and assume a Bloch type evanescent wave solution as $\psi(x,y)=(c_1,c_2)^{T} e^{-\lambda |y|} e^{i \Delta k_x x}$ with $\lambda>$0, one can get the eigenvector around boundary $y=0$
\begin{equation}
    \psi=\frac{1}{\sqrt{2}} (1,-1)^{T} e^{i \Delta k_x x} e^{-\lambda \left| y \right|},
\end{equation}
which yields the displacement
\begin{align}
    \bm{u}(\bm{k})=&( \bm{u}_1(\tau \bm{K})e^{i\tau \bm{K} \cdot \bm{r}},\bm{u}_2(\tau \bm{K})e^{i \tau \bm{K}\cdot \bm{r}})\cdot \psi\\
    =&\frac{1}{2}(i\tau,1,i \tau,-1) e^{i \tau \bm{K}\cdot \bm{r}} e^{i\Delta k_x x}e^{-\lambda \left| y \right|},
\end{align}
}
\textcolor{black}{so that the 
\begin{align}
    \binom{u_{px}}{u_{py}} =&\frac{1}{2} \binom{i \tau}{1}  e^{i\Delta k_x x}e^{-\lambda \left| y \right|}\\
    \binom{u_{qx}}{u_{qy}} =&\frac{1}{2} \binom{i \tau}{-1}  e^{i\Delta k_x x}e^{-\lambda \left| y \right|}.
\end{align} 
Based on the definition of $S_z$=$\frac{\rho\omega}{2}\text{Im}(u_x^*u_y-u_xu_y^*)$, we can obtain the SAM density at $p$ and $q$ sites as
\begin{align}
    S_{pz} =& \frac{\rho\omega}{2}\text{Im}(u_{px}^*u_{py}-u_{qx}u_{py}^*)=-\frac{\rho \omega \tau}{4} e^{-2\lambda \left| y \right|}\\
    S_{qz} =& \frac{\rho\omega}{2}\text{Im}(u_{qx}^*u_{qy}-u_{qx}u_{qy}^*)=\frac{\rho \omega \tau}{4} e^{-2\lambda \left| y \right|}
\end{align} 
}
\textcolor{black}{
Here we only briefly demonstrate the eigen states of this mass-spring model in connection with our continuum structures. Detailed theory can be found in Ref.~\cite{HGK2019JMPS} with a similar and fine mass-spring model.}

\textcolor{black}{According to above equation, at the interface between the structures with converse topological phases (see Supplementary Fig. \textbf{\ref{figs10}}), both $p$ and $q$ sites magnify non-zero vibration states, but the circularly polarization direction is opposite between $p$ and $q$ sites. Note that the sign of $\tau$ determines the propagative direction and circularly polarization direction of $p$ and $q$ points. 
When the propagation direction is reversed, the circularly polarization direction of $p$ and $q$ sites is also reversed. These behaviors of polarization direction are same as the SAM direction of $p$ and $q$ points in the 3D continuum structures.}

\begin{figure*}[!htb]
  \centering
  \includegraphics[width=\linewidth]{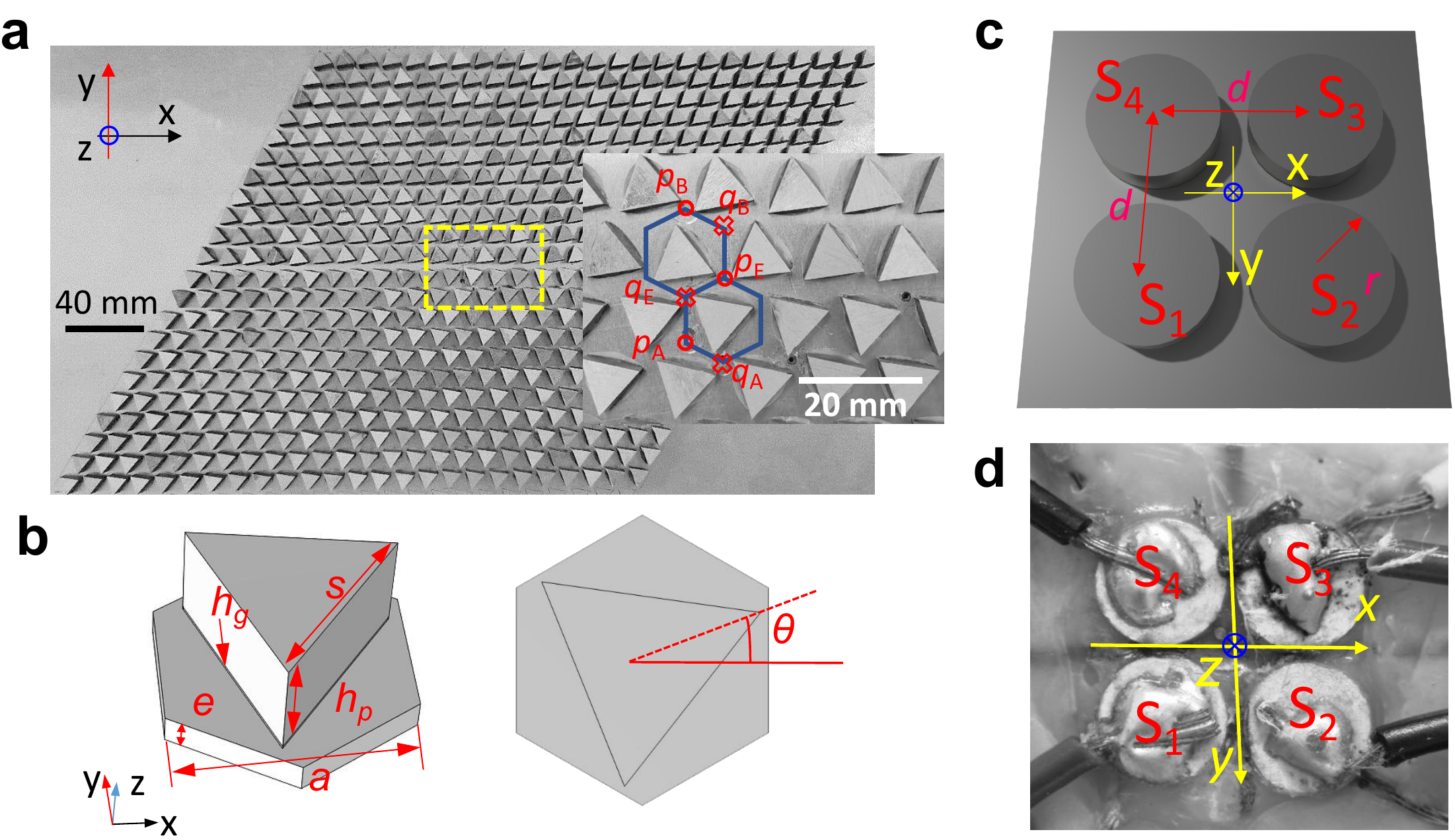}
  \caption{\textcolor{black}{\textbf{Topological valley PC plate and chiral elastic spin meta-source.} \textbf{a} Photograph of the PC plate fabricated by erecting triangular aluminum pillars in hexagonal lattice on an aluminum substrate. (Inset) The zoom in of the rectangular area inside yellow dashed box on the topological edge. The high symmetry points in each lattice are labeled as $p/q$ in turn. \textbf{b} Schematic of unit cell with the geometrical parameters including lattice constant $a$ = 12 mm, thickness of substrate plate $e$ = 1.5 mm, thickness of epoxy layer $h_g$ = 0.14 mm, the height, side length, and the rotation angle of triangular pillar $h_p$ = 5 mm, $s$ = 10 mm, $\theta$ = $\pm22^{\circ}$ respectively.  Schematic (\textbf{c}) and photograph (\textbf{d}) of bottom view of chiral elastic spin meta-source consisting four small PZT disks, S1-S4, arranged in square with distance $d$ = 6 mm. The PZT disks are made of PZT-5H with thickness 1.25 mm and radii $r$ = 2.5 mm. Four independent electrical signals are loaded on S1, S2, S3 and S4 with the same amplitude but every $\pm\pi/2$ phase difference versus the central frequency of input signals. Although the assembly of the meta-source occupies area about 11 mm$\times$11 mm, however the effective center area of meta-source has size only about 2.5 mm$\times$2.5 mm. }
  }
\label{figs1}
\end{figure*}

\begin{figure*}[!htb]
  \centering
  \includegraphics[width=\linewidth]{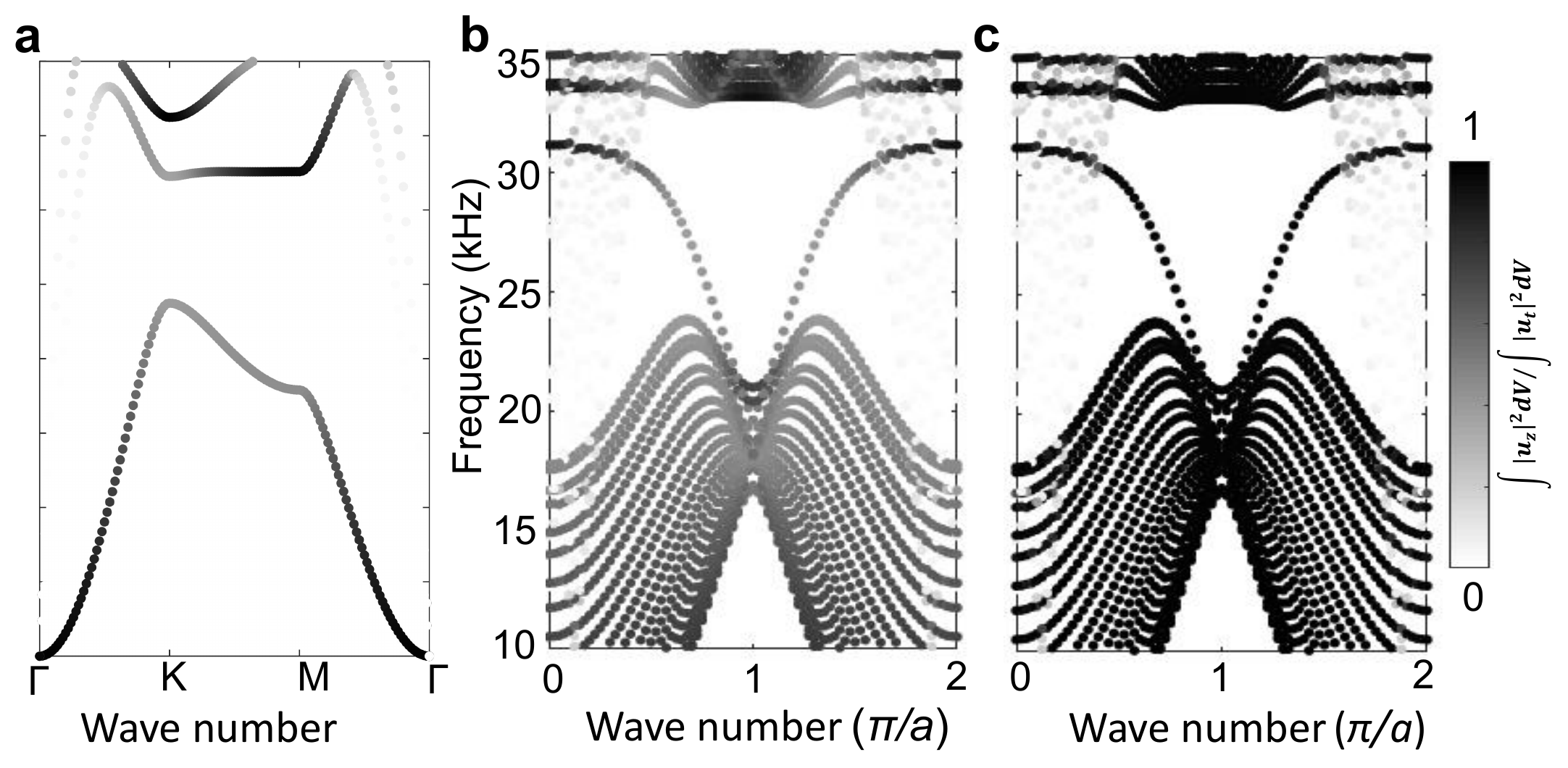}
  \caption{\textbf{Band structure of unit cell \textcolor{black}{and supercell} with $u_z$ proportion.} \textcolor{black}{\textbf{a} Band structure of unit cell  with rotation angle $\theta$ = $\pm22^{\circ}$.} The color scale indicates the proportion of out-ot-plane displacement $u_z$ against total displacement $u_t$ in unit cell as $\int{|u_z|^2dV}/\int{|u_t|^2dV}$. There is a band gap for anti-symmetric plate wave between 23.8 kHz and 32.2 kHz. \textcolor{black}{\textbf{b,c} The dispersion of supercell containing interface $\phi$ where phase B ($\theta$ = $-22^{\circ}$) up and phase A ($\theta$ = $22^{\circ}$) low, with the proportion of $u_z$ against total displacement $u_t$ in whole supercell (\textbf{b}) and only in substrate (\textbf{c}). Both of them show that the TES are dominated by the out-of-plane displacement.}}
\label{figs2}
\end{figure*}

\begin{figure*}[!htb]
  \centering
  \includegraphics[width=\linewidth]{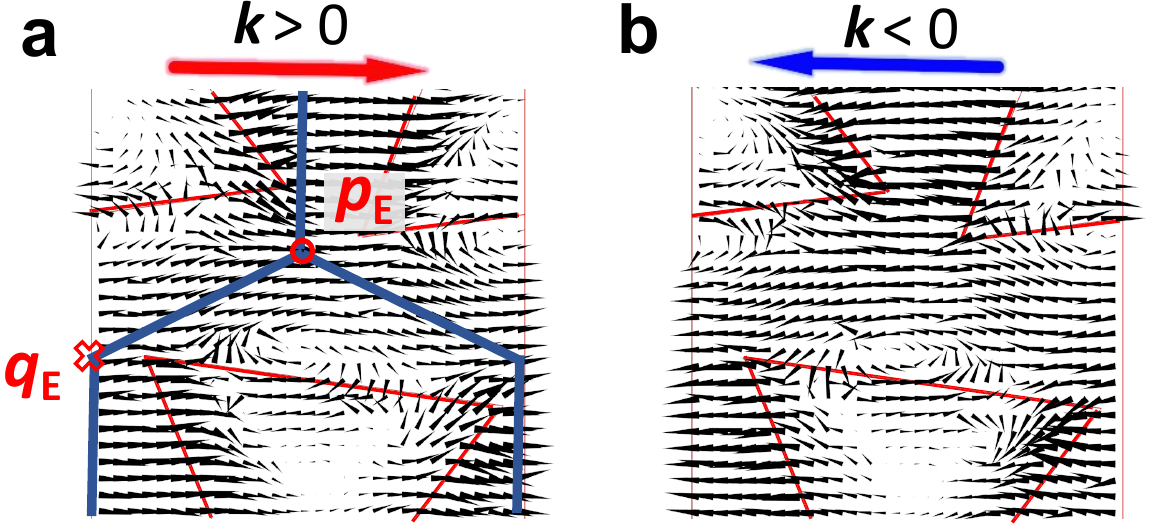}
  \caption{\textcolor{black}{The in-plane mechanical energy flux arrows of states with positive ($\pmb{k}>0$) and negative ($\pmb{k}<0$) group velocity at 25.8 kHz in confined areas in main Fig. \textbf{1b}.The energy flux has vortex somewhere, but it does not maintain distribution as clear as the elastic SAM density. In particular, the direction of energy flux is nearly flat at both $p_E$ and $q_E$ points. Thus, it is difficult to tell the difference about chiral properties between $p_E$ and $q_E$ points by using the vortex of energy flux. Note that we only show the distribution of energy flux on substrate, considering the meta-source is erected on the substrate and then the spin texture of meta-source is directly coupled to the substrate.}}
\label{figs9}
\end{figure*}

\begin{figure*}[!htb]
  \centering
  \includegraphics[width=\linewidth]{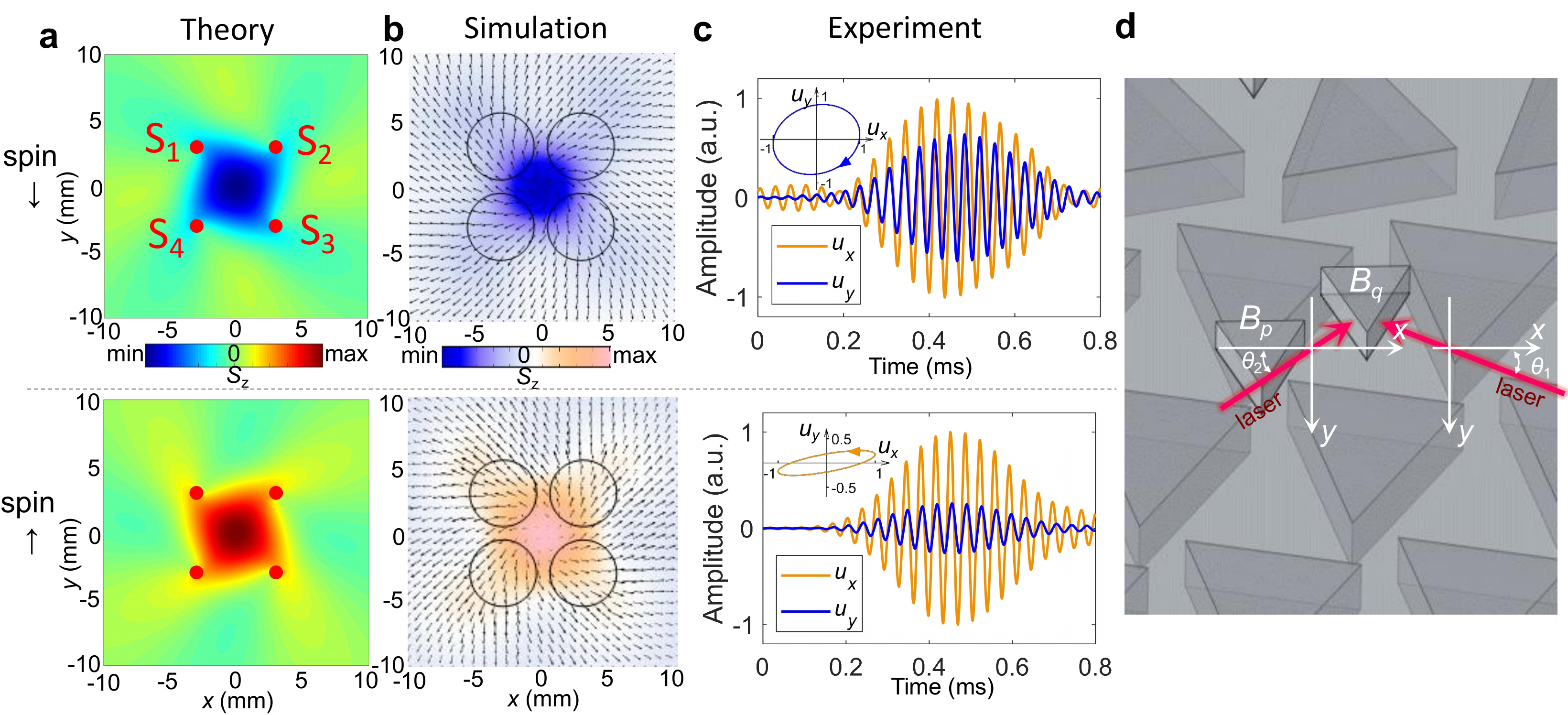}
  \caption{\textbf{Elastic spin of chiral elastic spin meta-source.} \textbf{a} Theoretically calculated $S_z$ (rainbow color table) with spin-down (spin $\downarrow$) or spin-up (spin $\uparrow$) excitation at 28 kHz on a 1.5 mm thick aluminum plate, whereby $S_z$ show negative or positive high density spot in central area. Red dots indicate the position of point sources. \textbf{b} Numerically calculated $S_z$ (twilight color table) and in-plane mechanical energy flux (black arrows) distribution for spin-down (spin $\downarrow$) or spin-up (spin $\uparrow$) source, whereby $S_z$ magnifies negative or positive high density spot in the center, as well as clockwise or anti-clockwise vortex energy flux.  Black circles indicate the position of sub-wavelength PZTs. \textbf{c} Measured time evolution of $u_x$ (orange line) and $u_y$ (blue line) at effective points at the center of meta-source on plate when excited with central frequency 28 kHz, and they are normalized to the maximum value of $u_x$ at each effective point. 
  Obviously, the $u_x$ is later than $u_y$ with with spin-down excitation. By extracting amplitude and phase information from displacement signals at given frequency (28 kHz here), the displacement polarization $\pmb{u}$ = ($u_x$, $u_y$) and the elastic $S_z$ can be obtained. The corresponding displacement polarization is clockwise and the $S_z$ is negative. 
  The phase of $u_x$ is earlier than $u_y$ when with spin-up excitation, and relevant displacement polarization is anti-clockwise and the $S_z$ is positive. \textbf{d} Schematic diagram of experimental measurements of elastic SAM density anywhere of the PC plate (shown here) and blank aluminum plate. To measure $S_z$, we put subwavelength triangular aluminum pillars on the back surface of substrate plate. Then we use laser Doppler vibrometer (LDV) to record the out-of-plane displacement versus two side faces of triangular pillar. Because the triangular pillar has relatively much small size, i.e. side length 2.9 mm and height 3 mm, the measured displacement can be seen at an effective point inside the pillar. The $S_z$ distribution inside triangular pillar has the same sign with corresponding point at our working frequency range. Here we label the displacement on $+x$ side as $u_1$ with angle $\theta_1$, while the displacement on $-x$ side as $u_2$ with angle $\theta_2$, so that  $u_x$ and $u_y$ of the effective point can be obtained as $u_x$ = $u_1 cos(\theta_1)$ - $u_2 cos(\theta_2)$, and $u_y$ = $u_1 sin(\theta_1)$ + $u_2 sin(\theta_2)$.
}
\label{figs3}
\end{figure*}

\begin{figure*}[!htb]
  \centering
  \includegraphics[width=\linewidth]{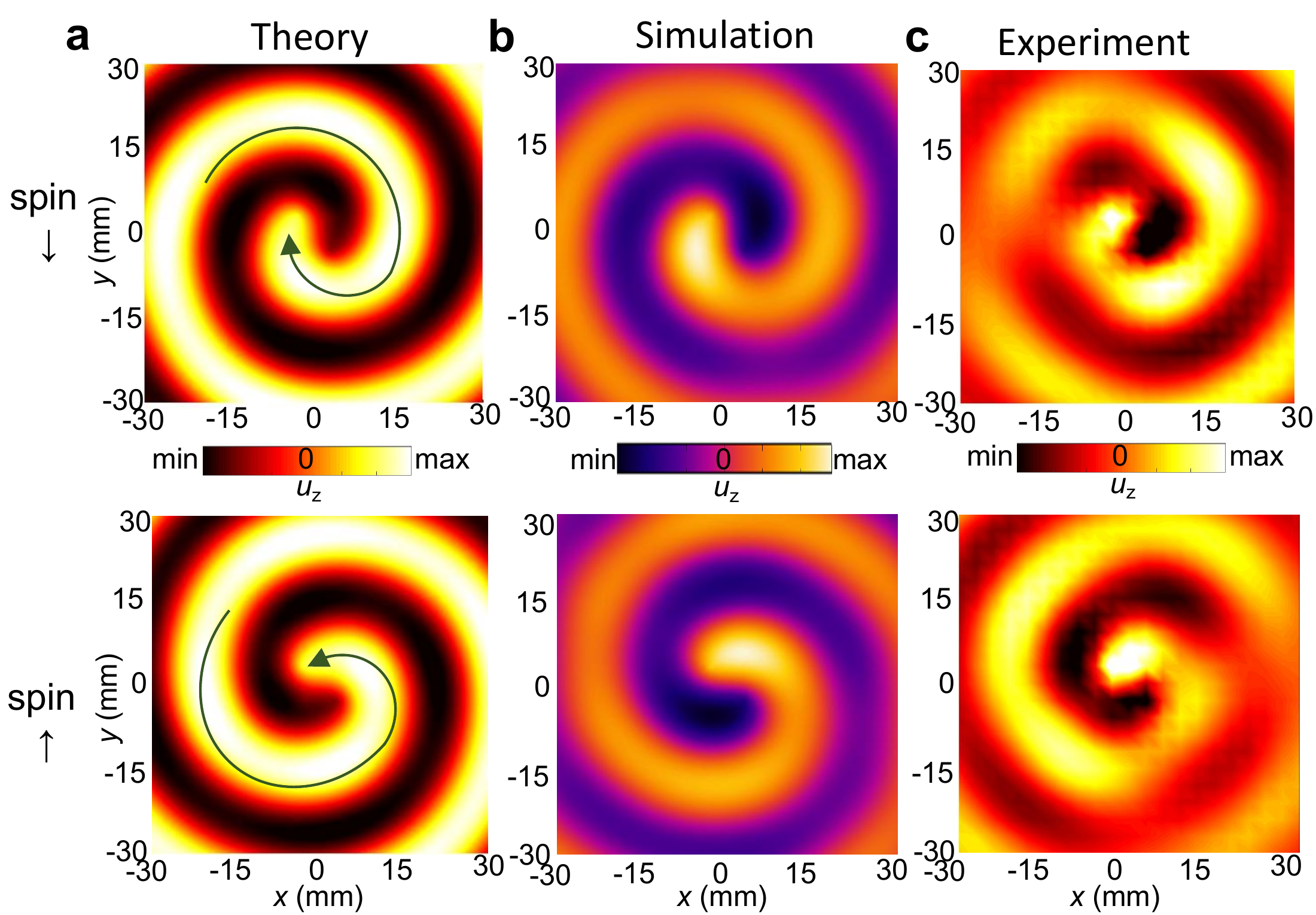}
  \caption{\textbf{Out-of-plane displacement $u_z$ by chiral elastic spin source.}  Theoretical \textbf{(a)}, numerical \textbf{(b)} and experimental \textbf{(c)} out-of-plane displacement $u_z$ distribution by chiral meta-source, all indicating clockwise (spin-down excitation) or anti-clockwise (spin-up excitation) rotation indicated by the arrows.
}
\label{figs4}
\end{figure*}


\begin{figure*}[!htb]
  \centering
  \includegraphics[width=\linewidth]{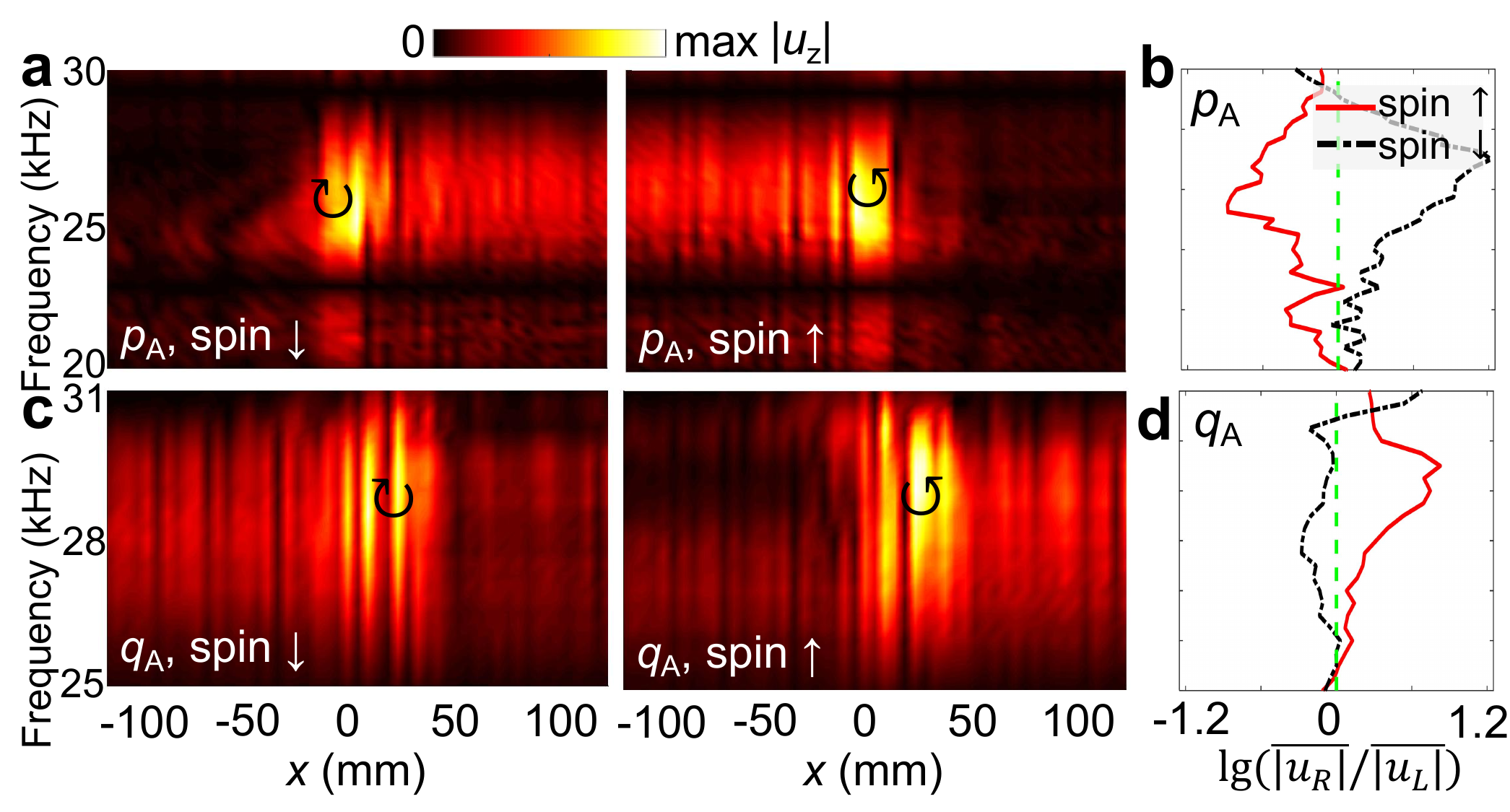}
  \caption{\textcolor{black}{\textbf{Experimental observation of unidirectional wave transportation when meta-source located around $p_A$ and $q_A$}. \textbf{a} The map of normalized $|u_z|$ measured along interface $\phi$ from $x$ = -125 mm to 125 mm every 5 mm  with chiral elastic spin meta-source centered at $p_A$, where $x$ = 0 is at the middle position along interface $\phi$. \textbf{b} The difference of $|u_z|$ between right and left side of meta-source, defined as lg$(\overline{|u_{R}|}/\overline{|u_{L}|})$, where $\overline{|u_{R}|}$ (resp. $\overline{|u_{L}|}$) is the averaged amplitude of $|u_z|$ measured at the right (resp. left) side of source, when meta-source is centered at $p_A$. The red solid and black dashed lines present the result with spin-up (spin $\uparrow$) and spin-down (spin $\downarrow$) sources, respectively. \textbf{c,d} The map of normalized $|u_z|$ (\textbf{c}) and difference (\textbf{d}) measured along interface $\phi$ with meta-source locating at $q_A$. In panel \textbf{a}, the asymmetric rightward and leftward wave propagation are successfully generated by spin-down and spin-up source, respectively. In panel \textbf{b}, the asymmetric behavior is clearly observed in broad frequency range where the positive (resp. negative) value indicates the rightward (resp. leftward) wave propagation, reaching a huge difference of $15.8$ times (\textcolor{black}{lg$(\overline{|u_{R}|}/\overline{|u_{L}|})$}$=1.2$). While in panels \textbf{c} and \textbf{d}, the situations are reversed: the spin-down and spin-up source generate the asymmetric leftward and rightward wave propagation, respectively, instead. We note that the relatively narrower frequency range between 25 kHz$\sim$31 kHz in panel \textbf{d} shall come from experimental factors including the the imperfect installation of meta-source and the sample parameter fluctuation.
  }}
\label{figs5}
\end{figure*}

\begin{figure*}[!htb]
  \centering
  \includegraphics[width=\linewidth]{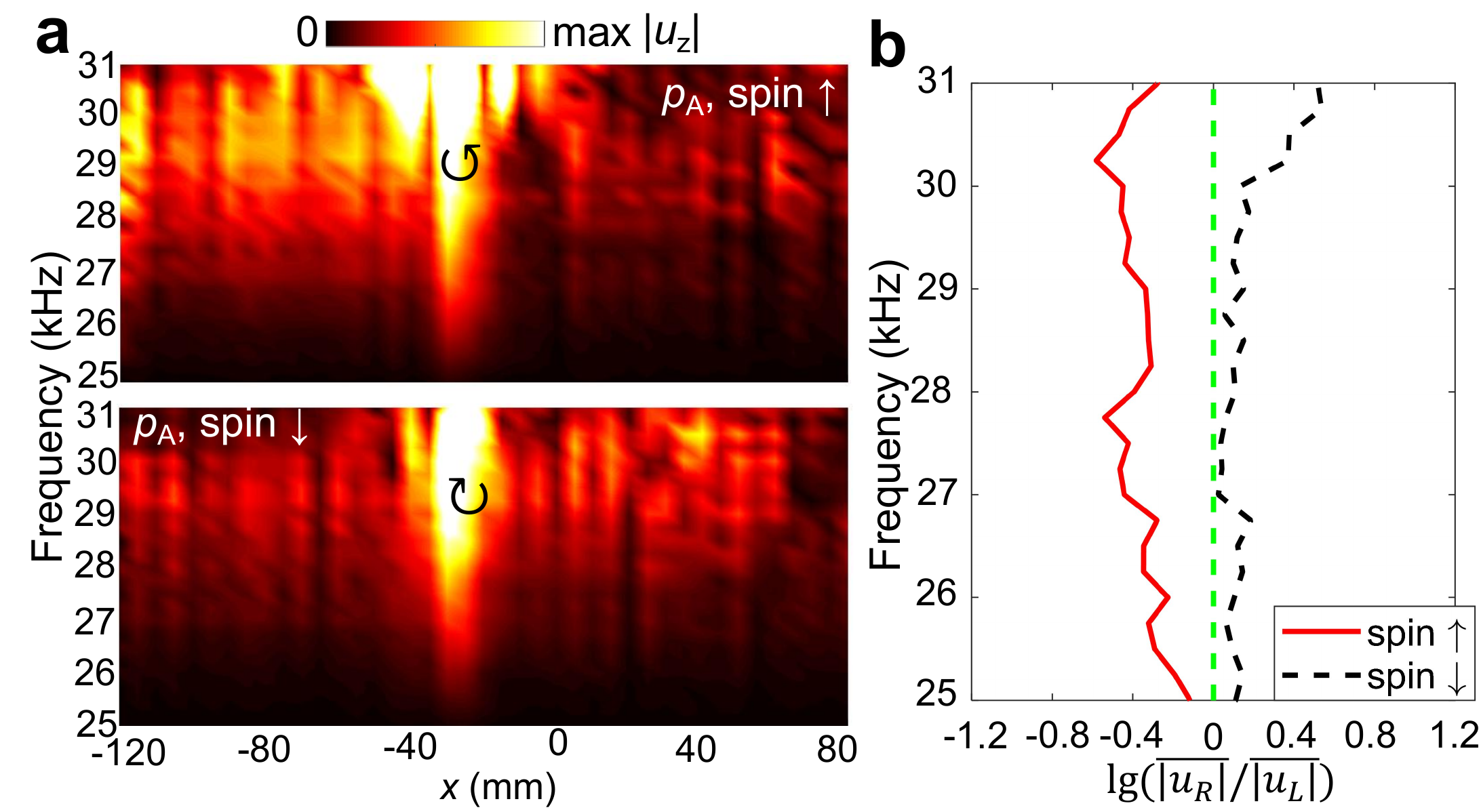}
  \caption{\textcolor{black}{\textbf{Experimental observation of unidirectional wave transportation by an array of four PZT disks with radii $r$ = 1.5 mm and distance $d$ = 4 mm}. \textbf{a} The map of normalized $|u_z|$ measured along interface $\phi$ from $x$ = -120 mm to 80 mm every 5 mm with chiral elastic spin meta-source centered at $p_A$. Here, $x$ = 0 is at the middle position along interface $\phi$, and the meta-source is installed at $x$ $\approx$ -25 mm. \textbf{b} The difference of $|u_z|$ between right and left sides of meta-source, defined as lg$(\overline{|u_{R}|}/\overline{|u_{L}|})$, where $\overline{|u_{R}|}$ (resp. $\overline{|u_{L}|}$) is the averaged amplitude of $|u_z|$ measured at the right (resp. left) side of source. The red solid and black dashed lines present the result with spin-up (spin $\uparrow$) and spin-down (spin $\downarrow$) sources, respectively. When using spin-up (resp. spin-down) source, the leftward (resp. rightward) elastic wave is successfully generated. However, the difference is smaller than its counterparts when using larger meat-source. Here the assembly of the meta-source occupies area about 7 mm$\times$7 mm, and the effective center area of meta-source has size only about 1.9 mm$\times$1.9 mm.
  }}
\label{figs6}
\end{figure*}

\begin{figure*}[!htb]
  \centering
  \includegraphics[width=\linewidth]{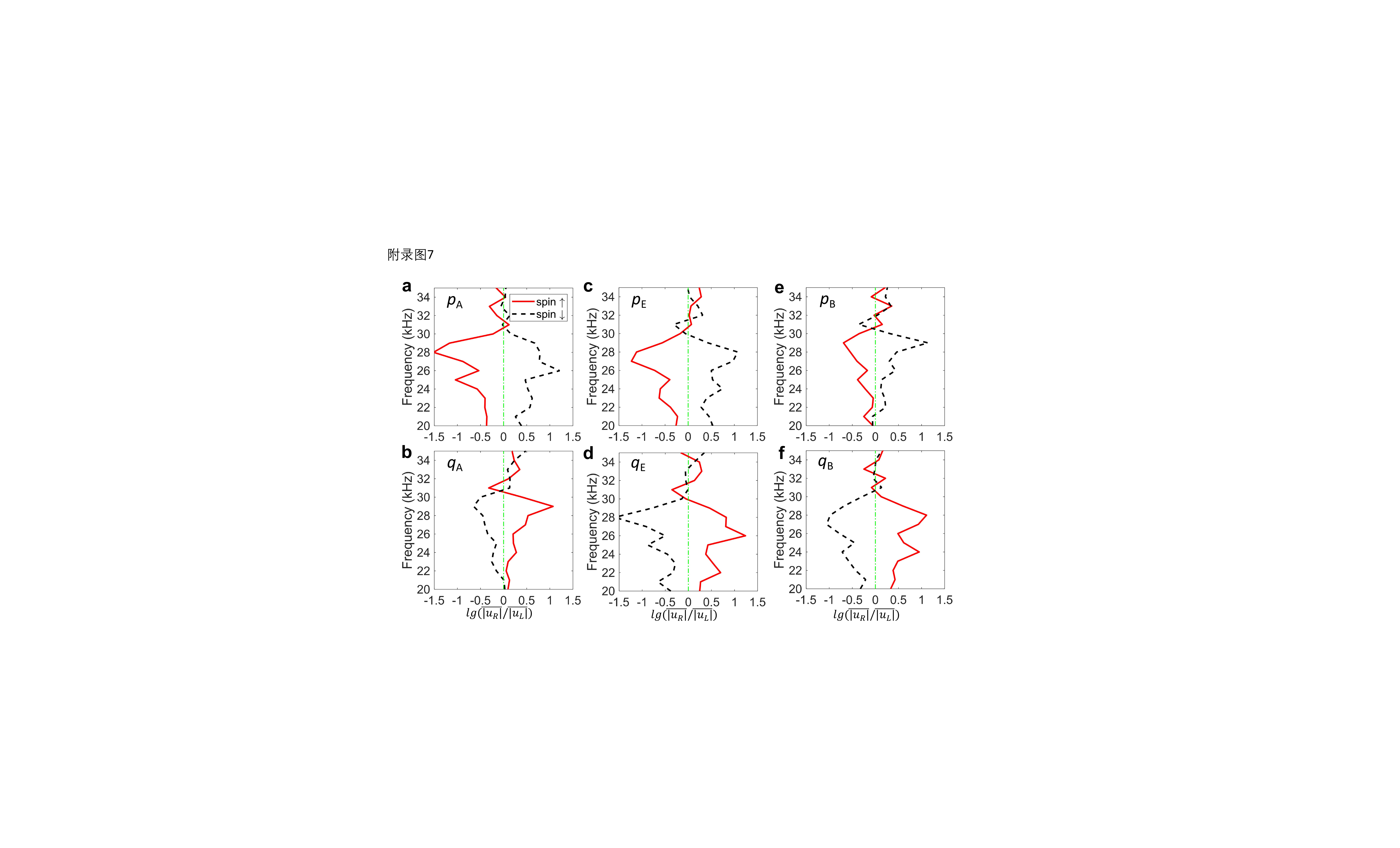}
  \caption{\textbf{The simulated difference of $|u_z|$ between right and left side of meta-source.} The average difference of $u_z$ between right and left side of meta-source centered at $p_A$ (\textbf{a}), $q_A$ (\textbf{b}), $p_E$ (\textbf{c}), $q_E$ (\textbf{d}), $p_B$ (\textbf{e}), and $q_B$ (\textbf{f}), respectively. The positive (resp. negative) value indicates the rightward (resp. leftward) transportation. The unidirectional propagation is clearly observed in a broad frequency range, and the difference can be up to 31 times \textcolor{black}{(lg$(\overline{|u_{right}|}/\overline{|u_{left}|})=1.5$)}. Obviously, the plate wave transports to left side with spin-up source but to right side with spin-down source all centered at $p$ points. On the other hand, the plate wave are rightward going with spin-up source but leftward going with spin-down source all centered at $q$ points, just opposite to those at $p$ points. The red solid (resp. black dashed) line presents the results with spin-up (resp. spin-down) source.}
\label{figs7}
\end{figure*}

\begin{figure*}[!htb]
  \centering
  \includegraphics[width=\linewidth]{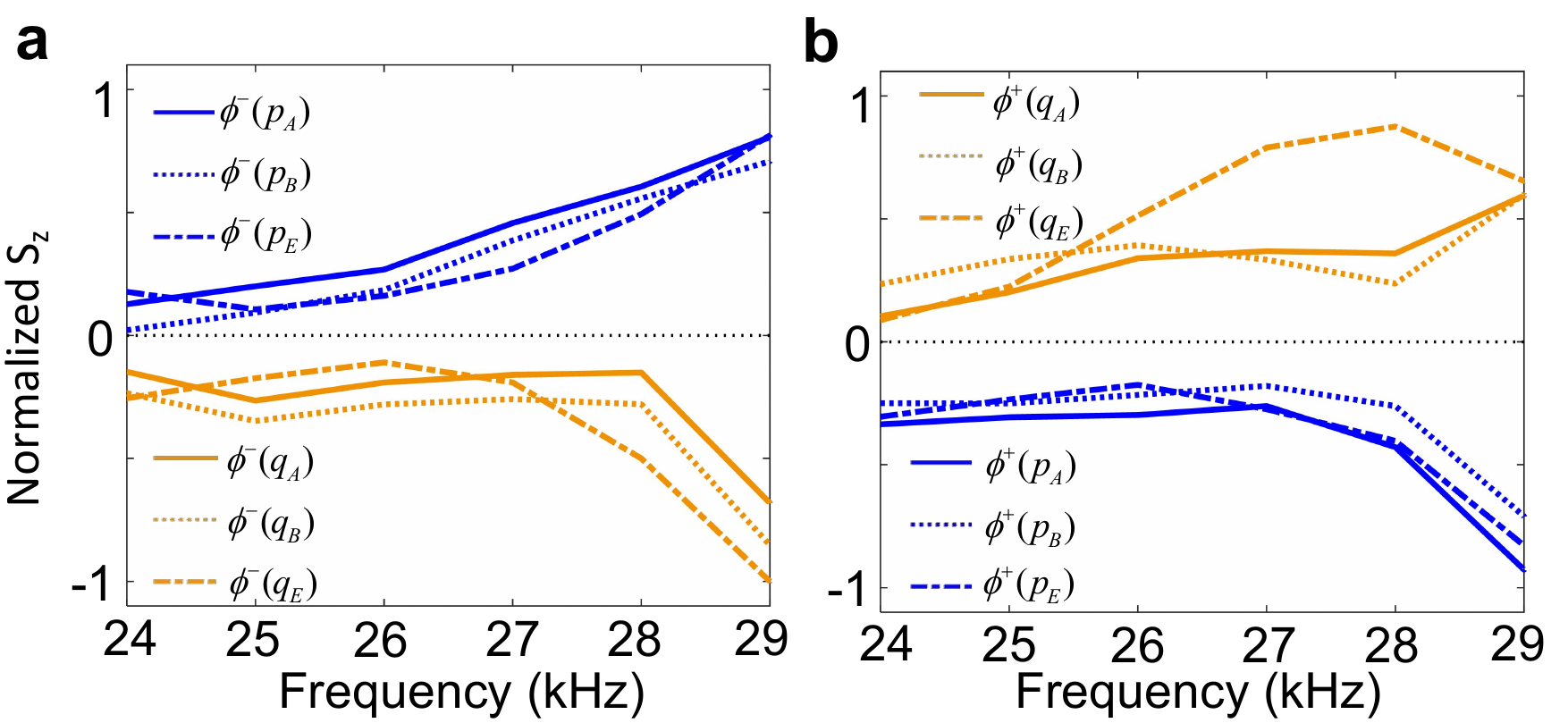}
  \caption{\textbf{Numerically simulated elastic $S_z$ at effective $p/q$ points.} \textbf{a,b} Computed $S_z$ \textcolor{black}{at left} (\textbf{a}) and \textcolor{black}{at right} (\textbf{b}) \textcolor{black}{side of point source} normalized to the overall maximum value in each panel. At \textcolor{black}{left side} and in a broad frequency range, $S_z$ features positive value on \textcolor{black}{$p$ points} but negative value on \textcolor{black}{$q$ points}. The situation at \textcolor{black}{right side} is reversed: $S_z$ is negative on \textcolor{black}{$p$ points} but positive on \textcolor{black}{$q$ points}.  
  }
\label{figs8}
\end{figure*}

\begin{figure*}[!htb]
  \centering
  \includegraphics[width=\linewidth]{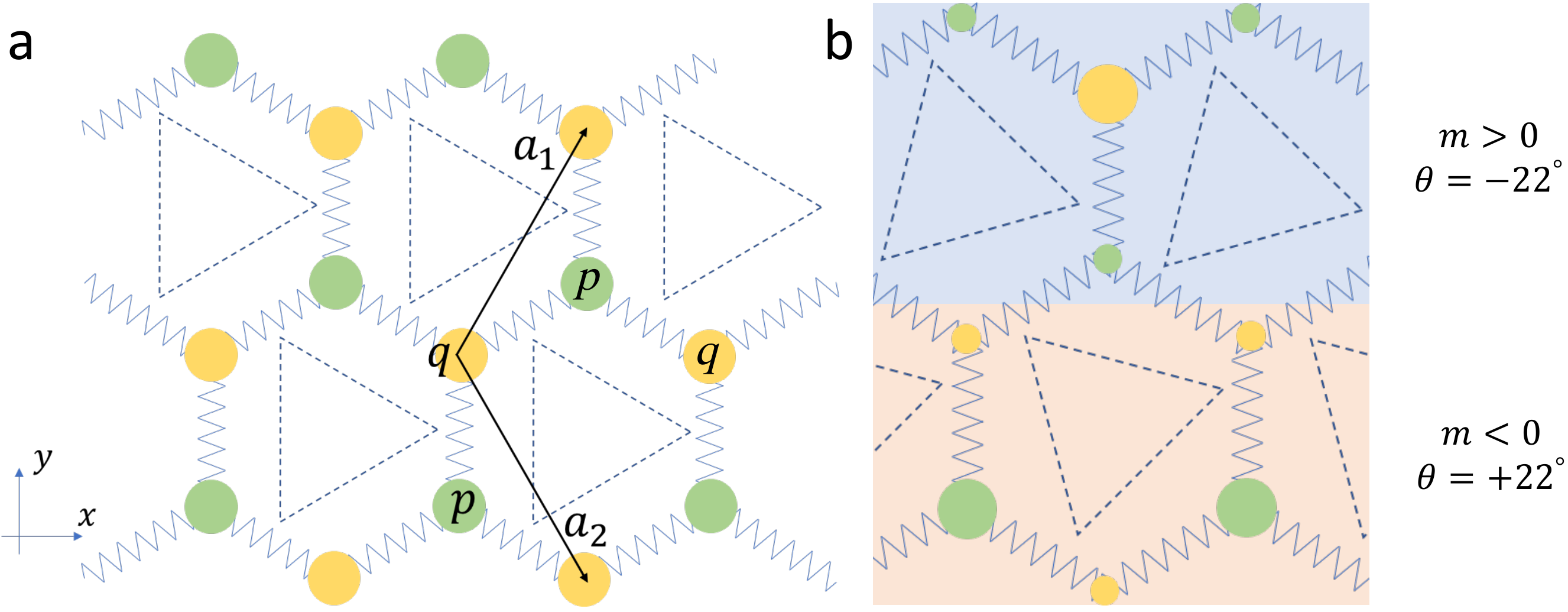}
  \caption{\textcolor{black}{\textbf{The mass-spring model.} \textbf{a} The orange circles are $p$ sites and blue circles are $q$ sites. Here, we use the black dashed lines to mark the triangular pillar in relevant valley metamaterial. \textbf{b} The boundary between PC with effective mass $m>0$ (up domain) and $m<0$ (bottom domain).}}
\label{figs10}
\end{figure*}

\end{document}